\documentclass[journal]{IEEEtran}
\usepackage{cite}
\usepackage{amsmath,amssymb,amsfonts}
\usepackage{algorithmic}
\usepackage{graphicx}
\usepackage{textcomp}
\usepackage{newunicodechar}
\newunicodechar{★}{\textasteriskcentered}
\usepackage{soul}
\usepackage{float}

\usepackage[caption=false,font=footnotesize]{subfig}
\usepackage{multirow}
\usepackage{array}
\usepackage{tabularx}
\usepackage[table]{xcolor}
\usepackage{braket}
\usepackage[hidelinks]{hyperref}

\def\BibTeX{{\rm B\kern-.05em{\sc i\kern-.025em b}\kern-.08em
    T\kern-.1667em\lower.7ex\hbox{E}\kern-.125emX}}

\begin{document}

\title{Indoor Occupancy Classification using a Compact Hybrid Quantum–Classical Model Enabled by a Physics-Informed Radar Digital Twin}

\author{Sebastian~Ratto~V.,~\IEEEmembership{Student Member,~IEEE,}
Ahmed~N.~Sayed,~\IEEEmembership{Senior Member,~IEEE,}
Neda~Rojhani,~\IEEEmembership{Senior Member,~IEEE,}
Arien~P.~Sligar,~\IEEEmembership{Member,~IEEE,}
Jose~R.~Rosas\mbox{-}Bustos,
Saasha~Joshi,
Luke~C.~G.~Govia,
Omar~M.~Ramahi,~\IEEEmembership{Fellow,~IEEE,}
and~George~Shaker,~\IEEEmembership{Senior Member,~IEEE}
\thanks{S. Ratto V., A. N. Sayed, N. Rojhani, O. M. Ramahi, and G. Shaker are with the Department of Electrical and Computer Engineering, University of Waterloo, Waterloo, ON N2L 3G1, Canada (e-mail: srattovalderrama@uwaterloo.ca; ansayed@uwaterloo.ca; nrojhani@uwaterloo.ca; oramahi@uwaterloo.ca; gshaker@uwaterloo.ca).}
\thanks{A. P. Sligar is with Synopsys Inc., Portland, OR, USA (e-mail: Arien.Sligar@ansys.com).}
\thanks{J. R. Rosas-Bustos is with Applied Quantum Technologies (AQT), Waterloo, ON, Canada (e-mail: me@joserosas.org).}
\thanks{S. Joshi is with CMC Microsystems, Toronto, ON, Canada (e-mail: saasha.joshi@cmc.ca).}
\thanks{L. C. G. Govia is with CMC Microsystems, Waterloo, ON, Canada (e-mail: luke.govia@cmc.ca).}
\thanks{Corresponding author: Sebastian Ratto V.}}

\markboth{Ratto~V. \MakeLowercase{\textit{et~al.}}: Compact HQNN for Radar Occupancy via Physics-Informed Digital Twin}%
{Ratto~V. \MakeLowercase{\textit{et~al.}}: Compact HQNN for Radar Occupancy via Physics-Informed Digital Twin}

\maketitle

\begin{abstract}
Indoor occupancy classification enables privacy-preserving monitoring in settings such as remote elder care, where presence information helps triage alarms without cameras or wearables. Radar suits this role by sensing motion through occlusions and in darkness. Modern deep-learning pipelines are the standard for interpreting radar returns effectively; however, they are often parameter-heavy and sensitive at low signal-to-noise ratios (SNR), motivating compact alternatives like Hybrid Quantum Neural Networks (HQNNs).
A two-qubit HQNN is benchmarked against convolutional neural networks (CNNs) using a physics-informed $60\,\mathrm{GHz}$ digital twin and real radar measurements under matched training protocols. In clean conditions, the HQNN achieves high accuracy ($99.7\%$ synthetic; $97.0\%$ real) with up to $170\times$ fewer parameters ($0.066\,\mathrm{M}$). Its parameter efficiency is shown to be structural, as an ablation of the parameterized quantum circuit (PQC) causes sharp performance drops on real data (to $68.5\%$ and $31.5\%$ for the control heads). A domain-dependent sensitivity emerges under additive-noise evaluation, where the HQNN begins recovery earlier in synthetic data while CNNs recover more steeply and peak higher on real measurements. In label-fraction ablations, CNNs prove more sample-efficient on real Range-Doppler Maps (RDMs), with the performance gap being most pronounced (at $50\%$ labels, BA $0.89$–$0.99$ vs.\ HQNN $0.75$). On synthetic data, this gap narrows significantly, largely vanishing by the $50\%$ label mark.
Overall, the HQNN's value lies in parameter efficiency and a compact inductive bias that shapes its distinct sensitivity profile; this work establishes a rigorous baseline for hybrid quantum models in privacy-preserving radar occupancy sensing.
\end{abstract}

\begin{IEEEkeywords}
Digital twin simulations, FMCW radar, health care applications, hybrid quantum--classical networks, quantum machine learning.
\end{IEEEkeywords}

\section{Introduction}
\IEEEPARstart{F}{alls} are a leading cause of injury and death among older adults. Globally, falls account for about 684{,}000 deaths each year, with adults over 60 at the greatest risk. In the United States, older adult falls led to nearly 3{,}000{,}000 emergency department visits and more than 38{,}000 deaths in 2021~\cite{who_falls_2021,cdc_falls_facts_2024}. Beyond detection, a central challenge is alarm management: systems that escalate every suspected fall regardless of context create nuisance alarms and caregiver fatigue. Studies report high rates of false or clinically insignificant alarms in clinical monitoring and bed-exit alerts, motivating the addition of people counting or presence information to triage whether the person is alone or already accompanied~\cite{wen2024_iot_bedexit_jmir,ahrq_psnet_alarm_fatigue_2023}.
Multiple sensing modalities have been explored for fall detection and people counting, each with trade-offs~\cite{newaz2023_fall_detection_review}. Wearables can be accurate but depend on user compliance. Vision using RGB or depth provides rich features but faces acceptance and privacy barriers in bedrooms and washrooms and can fail with poor lighting or occlusion~\cite{mujirishvili2023_video_aal_acceptance}. Thermal cameras reduce identity exposure but still capture imagery. Passive infrared and pressure sensors are simple, but lack fine detail. Wi-Fi channel-state methods use existing infrastructure but often fail to generalize to new homes, and some benchmarks can be optimistic due to subject overlap between train and test sets~\cite{acmcsur2025_wifi_sensing}.
Radar offers a practical balance for assisted living, it is contactless and privacy-preserving, works in any lighting, can operate through obscurants like smoke or steam, and supports people counting from range–Doppler and micro-Doppler features~\cite{zhang2025_mmwave_med_survey}. However, radar sensing faces issues regarding multipath, clutter, variable signal-to-noise ratio (SNR), and a scarcity of labeled data across diverse environments. These factors introduce domain shift that hurts model generalization, a challenge reported across presence detection, activity recognition, and gait analysis~\cite{Zhang2020DopplerNet,Kiani2024HybridDCNN,Han2020RadarActivity,Wang2024MIMOTracking}. This has motivated various domain adaptation strategies~\cite{Du2020GRSL,Lang2019GRSL,Xu2025SourceFreeRadar,Pinyoanuntapong2023GaitSADA}.
To enable controlled comparison and reduce the cost and safety risks of data collection, we use a configurable geometric simulator (hereafter referred to as a digital twin, DT) to generate synthetic radar data that mirrors planned sensor placements and representative motion scenarios. High-frequency asymptotic solvers such as shooting-and-bouncing rays capture complex geometric interactions, including multipath and occlusion~\cite{Chipengo2021HighFid,Elbadrawy2024MapCon,Sayed2024UAVMicrowaveMag,Antemratto,RattoACES}. Efficient data generation is also theoretically relevant, as data availability underpins links between computational and learning benefits. In practice, the simulator approximates rather than exactly reproduces a measurement setup, thus calibration details are addressed in the Methods section.

Hybrid quantum–classical neural networks (HQNNs) offer a promising alternative for signal classification in noisy conditions. Small parameterized quantum circuits enable compact models that are significantly more parameter-efficient than deep classical networks~\cite{Tsang2023TQE}. Recent studies also suggest that quantum-informed classifiers can extract robust representations from noisy, real-world data~\cite{Grossi2022TQE,Krunic2022TQE}. While quantum approaches are now being evaluated for large-scale sensing tasks such as satellite imaging~\cite{Otgonbaatar2023TQE}, their utility in smaller, resource-constrained radar applications remains largely unexplored. This motivates our central question: can a minimal HQNN, built around a two-qubit circuit, serve as an effective and parameter-efficient alternative to deep neural networks for radar-based occupancy detection?

We evaluate our lightweight HQNN against three established classical benchmarks for radar data: DopplerNet, a task-specific convolutional neural network (CNN); ResNet-18, a widely used deep CNN; and EfficientNet-B0, a highly parameter-efficient architecture. Our contributions are as follows:  
(i) we develop a configurable geometric-simulation pipeline that generates a clean synthetic dataset, mitigating the cost and ethical constraints of real-world collection for assistive-technology applications;  
(ii) we evaluate model sensitivity to additive noise, measuring performance degradation on both synthetic and real data; and (iii) we provide an empirical demonstration that a compact HQNN with only 66{,}000 parameters achieves competitive performance relative to deeper CNNs, particularly in challenging synthetic mid-SNR regimes. This highlights its architectural utility and parameter efficiency, with evaluations performed independently across synthetic and real domains.

The rest of the paper is organized as follows; Section~\ref{sec:setup} describes the simulation pipeline and the hybrid and classical models. Section~\ref{sec:results} presents performance under clean and noisy conditions. Section~\ref{sec:discussion} interprets the findings and discusses limitations, including sim-to-real considerations. In Section~\ref{sec:conclusion} we present our concluding remarks.


\section{Methods and Experimental Setup}
\label{sec:setup}

\subsection{Geometric Simulator Pipeline}

\begin{figure*}[!htbp]
  \centering
  \subfloat[]{\includegraphics[width=0.45\textwidth]{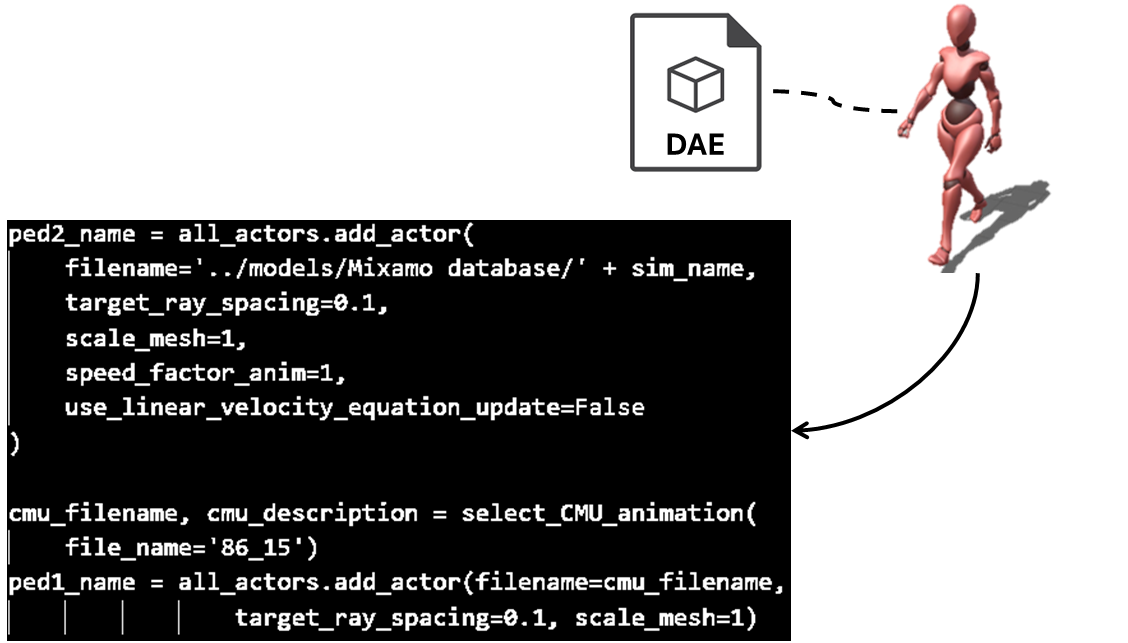}}
  \hfill
  \subfloat[]{\includegraphics[width=0.53\textwidth]{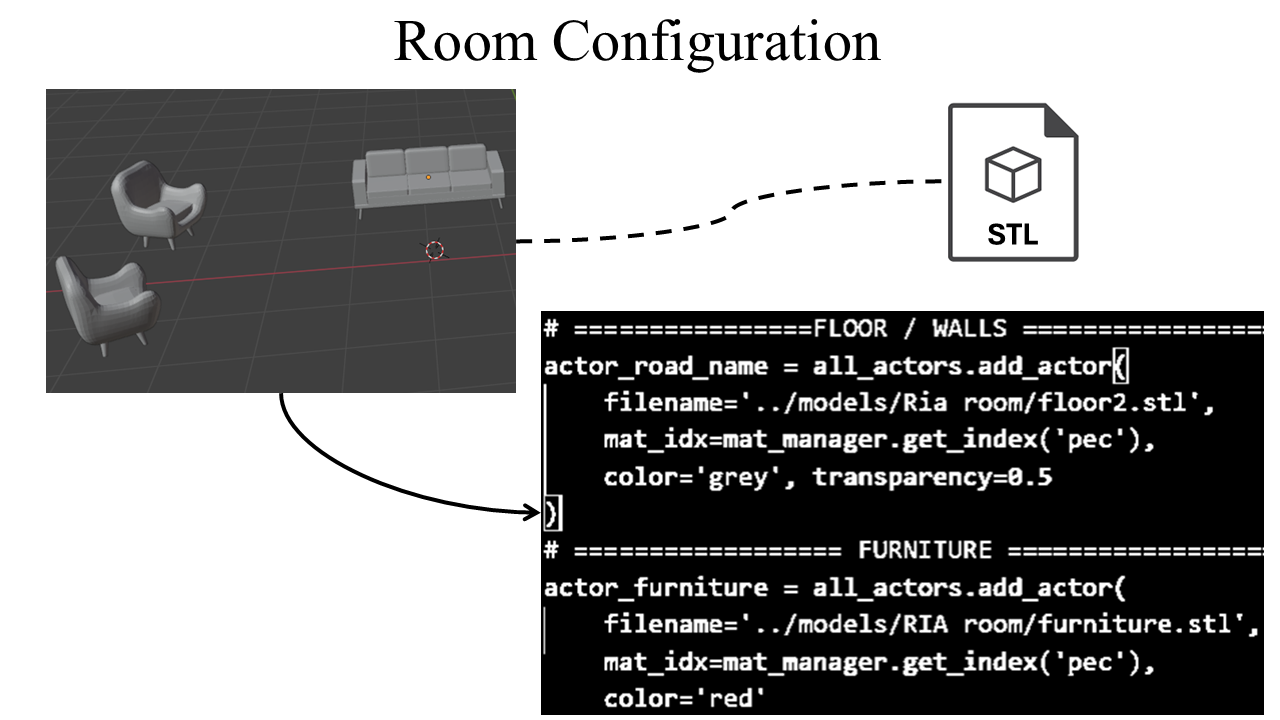}}
  \hfill
  \subfloat[]{\includegraphics[width=0.8\textwidth]{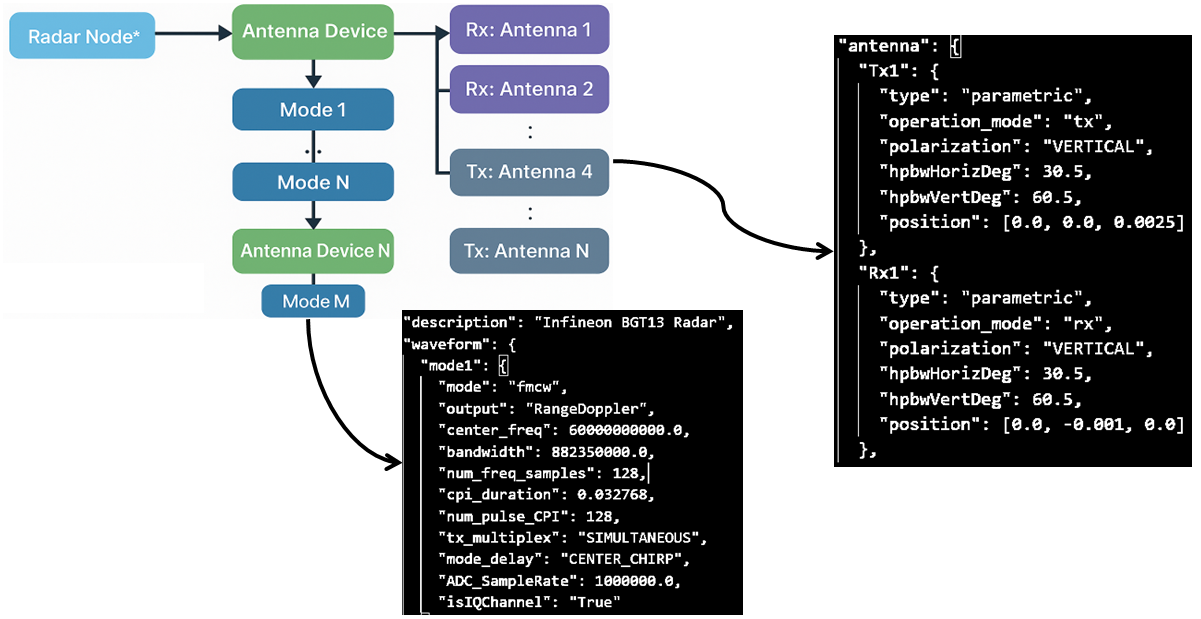}}

  \caption{Configuration modules of the digital-twin pipeline for FMCW radar simulation. 
  (a) Actor configuration: animated human models are imported using DAE files with parameterized motion, scale, and velocity. 
  (b) Room configuration: environment geometry is loaded from STL models with material assignments defining electromagnetic properties such as reflectivity and transparency. 
  (c) Sensor configuration: FMCW radar front-end, antenna placement, and waveform settings are specified to replicate the 60\,GHz device in simulation. 
  Together, these modules enable a reproducible and physically grounded digital-twin setup that captures human motion, clutter, and multipath propagation.}
  \label{fig:dt_config}
\end{figure*}

\begin{figure*}[!htbp]
  \centering
  \newlength{\panelH}\setlength{\panelH}{0.35\textheight}
  \newcommand{\panelgap}{12pt}

  \subfloat[]{\includegraphics[height=\panelH,keepaspectratio,width=\textwidth]{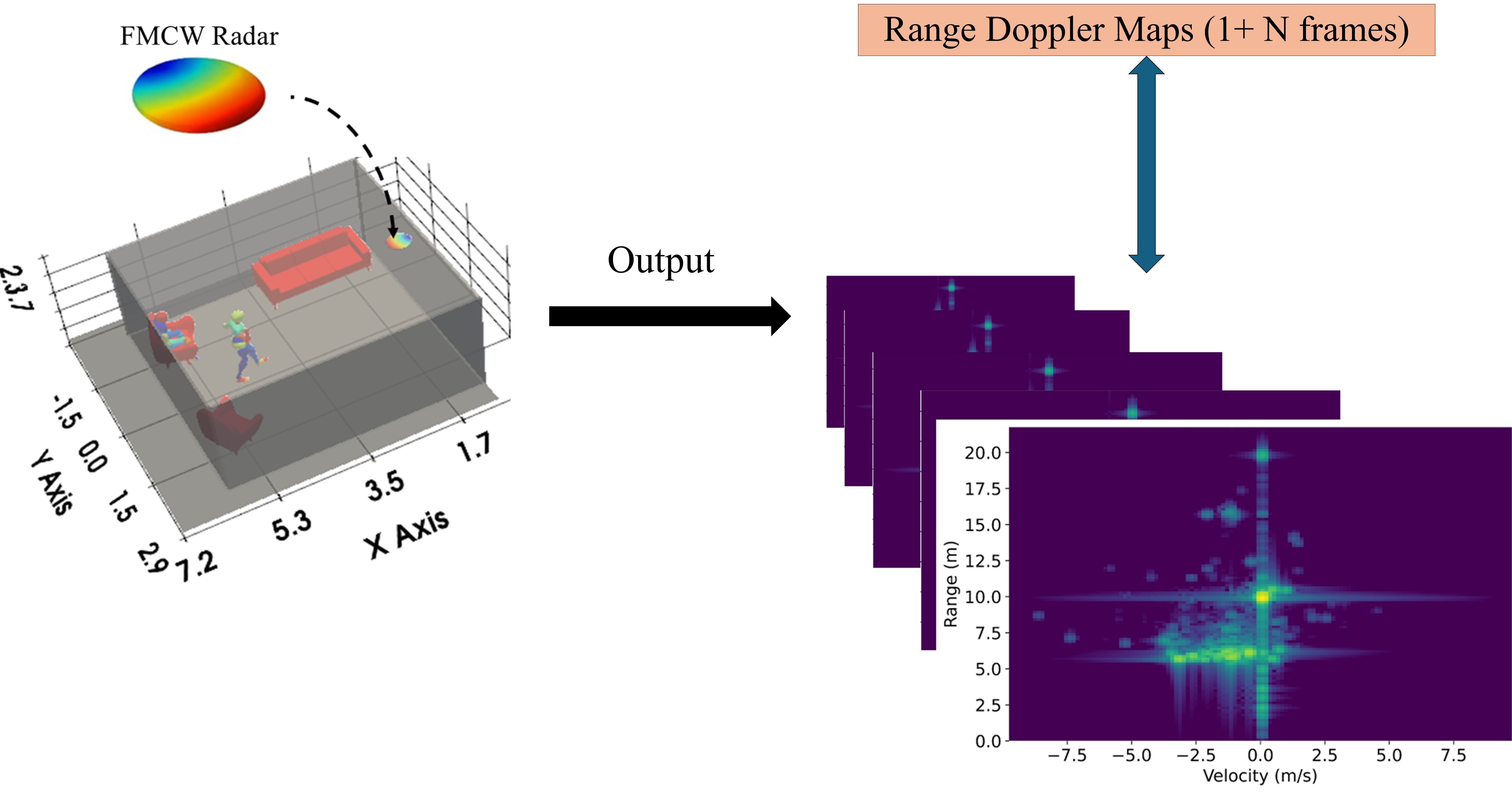}}\\[\panelgap]

  \subfloat[]{\includegraphics[height=\panelH,keepaspectratio,width=\textwidth]{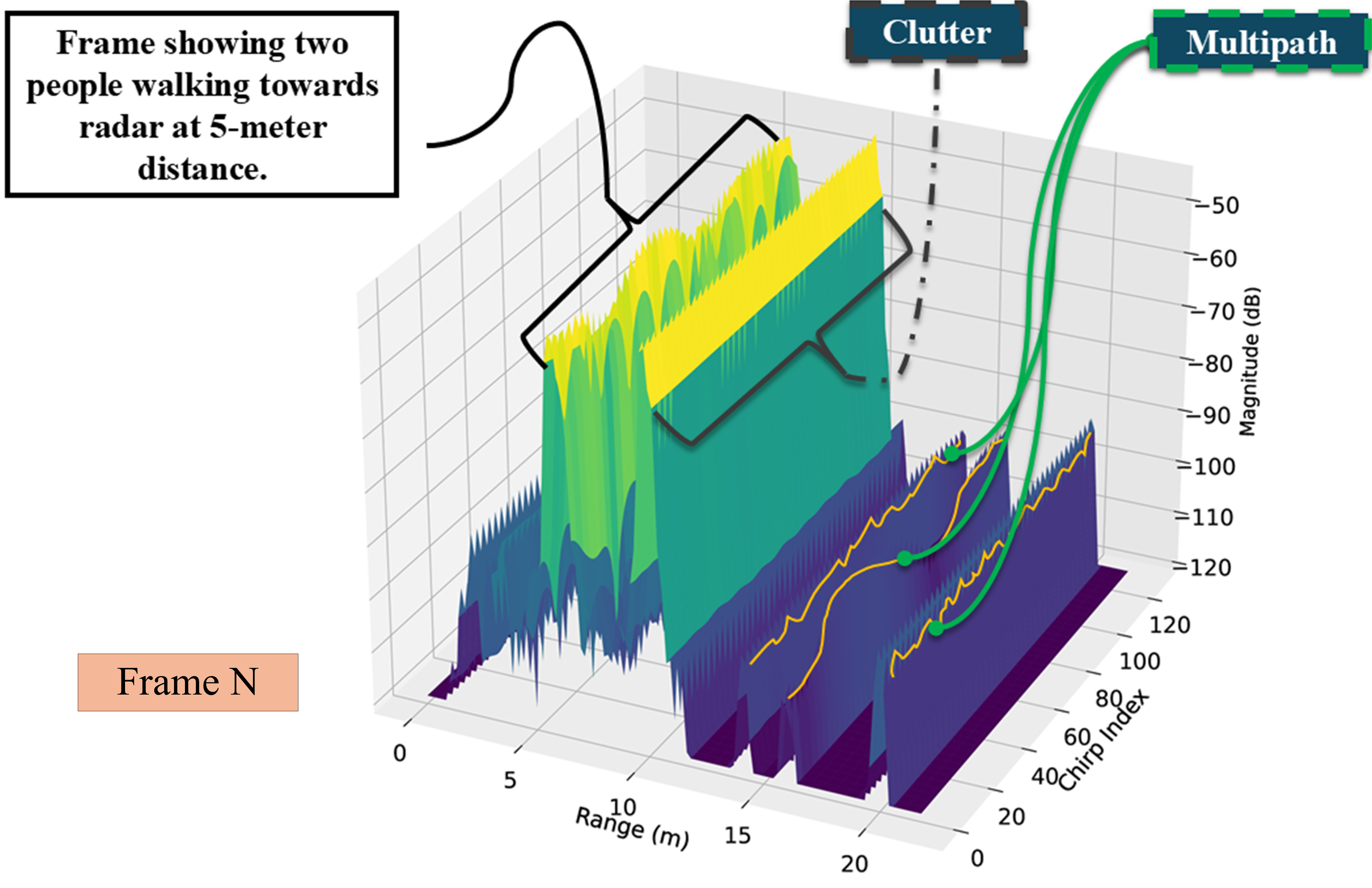}}

\caption{Digital-twin pipeline stages following scene configuration. 
(a) The simulation environment, defined programmatically, is executed and exported to a full-wave/high-frequency solver with assigned materials and radar parameters. The SBR method models multipath propagation, occlusions, and geometry-dependent interactions. (b) Radar signal processing converts the simulated baseband signals into RDMs, which form the dataset for training and evaluation of machine learning models.}

  \label{fig:dt_pipeline_cont}
\end{figure*}

A modular, code-driven geometric-simulation pipeline generates physics-informed Range Doppler Maps (RDMs) across varied layouts, placements, and occupancies. Animated motion and 3D scenes are authored in Blender~\cite{Blender2018}, using motions from Mixamo and the CMU Motion Capture Database to cover walking, squatting, turning, and entry/exit~\cite{Mixamo,CMU_Mocap}. Scenes are exported and simulated using the commercially available, physics-based electromagnetic solver, Synopsys Perceive~EM, where material properties and FMCW radar parameters are configured to match the BGT60TR13C module~\cite{InfineonBGT60TR13C}. An overview is shown in Fig.~\ref{fig:dt_config} and Fig.~\ref{fig:dt_pipeline_cont}.

All simulations are launched programmatically via a Python API to ensure reproducibility and scalable dataset synthesis. The workflow scripts scene import, radar placement/orientation, antenna patterns, sweep configuration, and batch exports, enabling systematic sweeps over layouts, materials, poses, and sensor placements. This produces labeled complex returns suitable for downstream signal processing and learning. 
Perceive~EM employs a \textit{Shooting-and-Bouncing Rays} (SBR) simulation methodology, based on a hybridization of \textit{Physical Optics} (PO) and \textit{Geometrical Optics} (GO) numerical techniques, with GPU acceleration to efficiently model electrically large environments~\cite{Chipengo2021HighFid}. 
The implementation incorporates diffraction effects, multipath scattering and transmission, and real material properties to generate high-fidelity synthetic backscatter data. This output integrates seamlessly into digital-twin pipelines via a lightweight API~\cite{ansys-perceive-em}. We refer to our setup as a geometric, physics-based simulator rather than a fully calibrated digital twin. A true DT would require absolute power/noise-floor calibration of the receiver chain and broader closed-loop sim-to-real checks, which are beyond the scope of this work. Accordingly, the simulated receiver chain is idealized, no receiver thermal noise or oscillator phase noise is modeled, and no Moving target indication (MTI), constant false alarm rate (CFAR) detection, or other background-suppression techniques are applied. Antenna patterns and placements mirror the BGT60TR13C datasheet and deployments, whereas materials use 60\,GHz EM properties for correct electromagnetic (EM) wave propagation modeling. Under these idealizations, the EM solution is the sole source of scene-dependent variability.

\subsubsection{SBR scattering model (GO/PO)}
Following Ling \emph{et~al.}~\cite{SBRcavity1991} (using the $e^{+j\omega t}$ convention), an incident plane wave is
\begin{IEEEeqnarray}{rCl}
\mathbf{E}^i(\mathbf{r})
&=& \bigl(-\hat{\phi}^{\,i}\, I \;+\; \hat{\theta}^{\,i}\, \bar{I}\bigr)
\, e^{j\,\mathbf{k}^{\,i}\!\cdot \mathbf{r}},
\label{eq:incident}
\end{IEEEeqnarray}
\noindent where $I$ and $\bar{I}$ are scalar amplitudes weighting the perpendicular
and parallel polarizations with respect to the plane of incidence (they are not unit vectors);
$\hat{\theta}^{\,i}$ and $\hat{\phi}^{\,i}$ are spherical unit vectors evaluated at the
incidence angles $(\theta^i,\phi^i)$; and
\begin{IEEEeqnarray}{rCl}
\mathbf{k}^{\,i}
&=& k_0\bigl(\hat{x}\sin\theta^{\,i}\cos\phi^{\,i}
            + \hat{y}\sin\theta^{\,i}\sin\phi^{\,i}
            + \hat{z}\cos\theta^{\,i}\bigr),
\label{eq:kvec}
\end{IEEEeqnarray}
with $k_0 = 2\pi/\lambda$.

Each ray is traced parametrically as
\begin{align}
(x_1, y_1, z_1) &= (x_0, y_0, z_0) + (s_x, s_y, s_z)\,t,
\label{eq:raytrace}
\end{align}
where $t$ is the path parameter, and the direction cosines are
\begin{IEEEeqnarray}{rCl}
s_x &=& -\sin\theta^i \cos\phi^i, \nonumber\\
s_y &=& -\sin\theta^i \sin\phi^i, \nonumber\\
s_z &=& -\cos\theta^i .
\label{eq:dircos}
\end{IEEEeqnarray}

At the aperture $\Sigma_A$ (the finite region of the $z=0$ plane), the outgoing field is replaced by an equivalent magnetic current sheet
\begin{equation}
\mathbf{K}_{\!s}(x,y)=
\begin{cases}
2\,\mathbf{E}(x_N,y_N,0)\times\hat{z}, & \mathbf{r}\in\Sigma_A,\\[2pt]
0, & \mathbf{r}\notin\Sigma_A,
\end{cases}
\label{eq:ksheet}
\end{equation}
where $(x_N,y_N,0)$ is the ray–aperture intersection and
$\mathbf{E}(x,y,0)=(E_x(x,y),\,E_y(x,y),\,0)$ is the tangential Cartesian electric field sampled on $\Sigma_A$, which serves as the integration surface for the equivalent current sheet and subsequent radiation integrals.

The monostatic backscattered far field is
\begin{align}
\mathbf{E}^s(r) &= \frac{e^{-j k_0 r}}{r}
    \Bigl( \hat{\theta}\, A_\theta^i + \hat{\phi}\, A_\phi^i \Bigr),
    && r \to \infty
\label{eq:scattered}
\end{align}
where \(A_\theta\) and \(A_\phi\) are scalar radiation coefficients evaluated
in the observation basis taken at the incidence direction $(\theta^i,\phi^i)$, and
$r$ is the distance from the aperture reference to the observation point.

\begin{IEEEeqnarray}{rCl}
A_\theta
&=& \frac{j k_0}{2\pi}
    \iint_{\Sigma_A}
    \bigl(\,E_x \cos\phi^i + E_y \sin\phi^i\,\bigr)
\nonumber\\
&& {}\qquad\cdot e^{j(k_x x + k_y y)}\, dx\,dy,
\label{eq:atheta}
\end{IEEEeqnarray}

\begin{IEEEeqnarray}{rCl}
A_\phi
&=& \frac{j k_0}{2\pi}
    \iint_{\Sigma_A}
    \bigl(\,-E_x \sin\phi^i + E_y \cos\phi^i\,\bigr)\cos\theta^i
\nonumber\\
&& {}\qquad\cdot e^{j(k_x x + k_y y)}\, dx\,dy .
\label{eq:aphi}
\end{IEEEeqnarray}
The transverse spectral wavenumbers appearing in the phase factor are
\begin{align}
k_x = k_0 \sin\theta^i \cos\phi^i, \qquad
k_y = k_0 \sin\theta^i \sin\phi^i ,
\label{eq:kxky}
\end{align}
\noindent Here $E_x(x,y)$ and $E_y(x,y)$ are the tangential Cartesian components of $\mathbf{E}(x,y,0)$ on the aperture $\Sigma_A$ (units V/m), and $k_x,k_y$ are the transverse spectral wavenumbers associated with the observation direction $(\theta^i,\phi^i)$; $k_0=2\pi/\lambda$.

\subsubsection{Radar model and signal chain}
The Infineon BGT60TR13C radar is emulated with carrier frequency 
$f_c=60$\,GHz and bandwidth $B=882.35$\,MHz. The sampling rate is set to 
$f_s=1$\,MHz and the chirp duration to $T_{\mathrm{chirp}}=0.256$\,ms, yielding 
$N=f_s T_{\mathrm{chirp}}=256$ samples per chirp. The chirp slope is
\begin{equation}
  \mu=\frac{B}{T_{\mathrm{chirp}}}=3.44\times 10^{12}\,\text{Hz/s}.
\end{equation}

Two fundamental limits follow from these parameters:
\begin{equation}
  \Delta R=\frac{c}{2B}=0.17\,\text{m}, \qquad
  R_{\max}=\frac{c\,f_s}{4\mu}=21.7\,\text{m},
\end{equation}
where $c$ is the speed of light. Here $\Delta R$ is the range resolution 
dictated by bandwidth, while $R_{\max}$ is the maximum unambiguous range imposed by the ADC sampling rate and chirp slope.

A linear FMCW chirp is modeled as
\begin{equation}
  s_{\mathrm{tx}}(t)=
  \exp\!\left\{j2\pi\!\left(f_c t+\tfrac{\mu}{2}t^2\right)\right\}.
\end{equation}
For a target at range $R$ and radial velocity $v$, the round-trip delay is 
$\tau=2R/c$ and the Doppler frequency is $f_D=2v/\lambda$, with 
$\lambda=c/f_c$ the wavelength. After mixing the received echo with 
$s_{\mathrm{tx}}(t)$ and low-pass filtering, the beat signal is
\begin{equation}
  s_b(t)\approx A\,
  \exp\!\left\{j2\pi \big((\mu\tau+f_D)t\big)\right\},
\end{equation}
where $A\in\mathbb{C}$ accounts for two-way propagation loss, antenna gains, 
target scattering (RCS), and a constant phase. The beat frequency is
\begin{equation}
  f_b \approx \mu\tau+f_D,
\end{equation}
which is approximately additive for an up-chirp and small $|f_D|$. 
The measured beat frequency maps to the per-target estimated range
\begin{equation}
  R \approx \frac{c}{2\mu}\,f_b,
\end{equation}
valid so long as $R<R_{\max}$.

With $M$ chirps and repetition interval $T_r$, Doppler processing yields
\begin{equation}
  v_{\max}=\frac{\lambda}{4T_r}, \qquad
  \Delta v=\frac{\lambda}{2MT_r},
\end{equation}
where $v_{\max}$ is the maximum unambiguous velocity and $\Delta v$ is the 
Doppler resolution.

A two-dimensional range--Doppler map (RDM) is then formed by applying a 
1-D FFT along fast time (samples within each chirp), followed by a 1-D FFT 
across chirps at each range bin. Denote $s[n,m]$ as the beat signal at 
fast-time index $n=0,\dots,N-1$ and chirp index $m=0,\dots,M-1$. The range FFT is
\begin{equation}
  S_r[k,m] = \sum_{n=0}^{N-1} s[n,m]\,
  e^{-j2\pi kn/N}, \qquad k=0,\dots,N-1,
\end{equation}
and the Doppler FFT produces the RDM:
\begin{IEEEeqnarray}{rCl}
  \mathrm{RDM}[k,\ell]
  &=& \Bigg|\sum_{m=0}^{M-1} S_r[k,m]\,
      e^{-j2\pi \ell m/M}\Bigg|^{2}, \nonumber\\
  && \ell=0,\dots,M-1.
\end{IEEEeqnarray}

Each cell $(k,\ell)$ corresponds to a discretized range--velocity bin. 
Raw complex fields from SBR simulations are passed through this ideal FMCW 
chain to generate 2-D RDMs, stored in dB units per frame. Unlike a standard image used in machine learning classification, each frame in the RDM represents measured signal energy values across range and Doppler bins. These values encode physical radar returns rather than visual pixel intensities, which means the "image-like" representation has a different statistical structure and interpretation.

\medskip
\noindent
The radar parameters are selected to balance hardware capabilities, sensing 
requirements, and computational efficiency. The carrier frequency is fixed at 
$f_c=60$\,GHz by the chipset, placing operation in the mmWave band and enabling 
fine spatial resolution. A target resolution of $\Delta R=0.17$\,m, sufficient 
to distinguish single-person occupancy, dictates the required bandwidth of 
$B=882.35$\,MHz. The sampling rate and chirp duration are configured to produce 
$N=256$ samples per chirp, aligning with hardware front-end constraints while 
ensuring FFT lengths that are computationally efficient for large-scale 
digital-twin experiments. A Hann window is applied before the FFT to suppress 
sidelobes and improve dynamic range. As a result, parameters such as the chirp 
slope $\mu$ and maximum unambiguous range $R_{\max}$ are not independent design 
choices but follow directly from these application-driven specifications. The 
sensing environment further influences radar placement, scattering, and clutter 
characteristics. In the corridor case (12\,m $\times$ 2\,m), the narrow geometry 
produces strong waveguiding effects, with wall reflections reinforcing or 
canceling the direct return. Multipath energy can persist over long distances, 
complicating separation of target motion from clutter. Placement of the radar at 
one end provides coverage across the corridor but also increases sensitivity to 
constructive and destructive interference patterns. In contrast, the room case 
(8\,m $\times$ 6\,m) introduces rich multipath from furniture, floor, and ceiling, 
where objects such as the sofa and chair act as scatterers producing clutter 
overlapping with the main Doppler responses. The shorter dimensions reduce 
long-range coherence of multipath but create localized hot-spots and blind zones 
depending on furniture arrangement.
Since the operating frequency is fixed at 60\,GHz ($\lambda \approx 5$\,mm), even 
small displacements or object edges on the order of a few centimeters produce significant scattering. These effects highlight the need
for environment-specific placement strategies and motivate digital twin simulations
that systematically capture multipath and clutter.

\subsubsection{Scenes, Placements, and Occupancy}
\label{sec:scenes}
\begin{figure}[!t]
\centering
\includegraphics[width=0.4\textwidth]{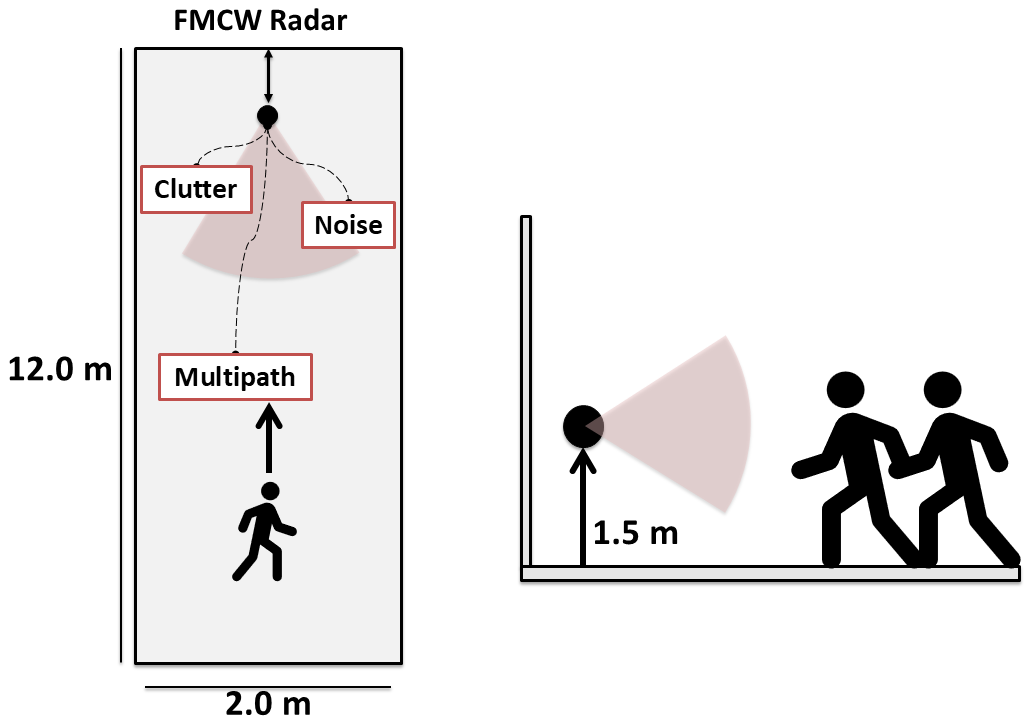}
\caption{Schematic layout of the simulated corridor environment (12\,m $\times$ 2\,m) showing radar placement, representative multipath, and clutter/noise regions.}
\label{fig:test_corridor}
\end{figure}

\begin{figure}[!t]
\centering
\includegraphics[width=0.45\textwidth]{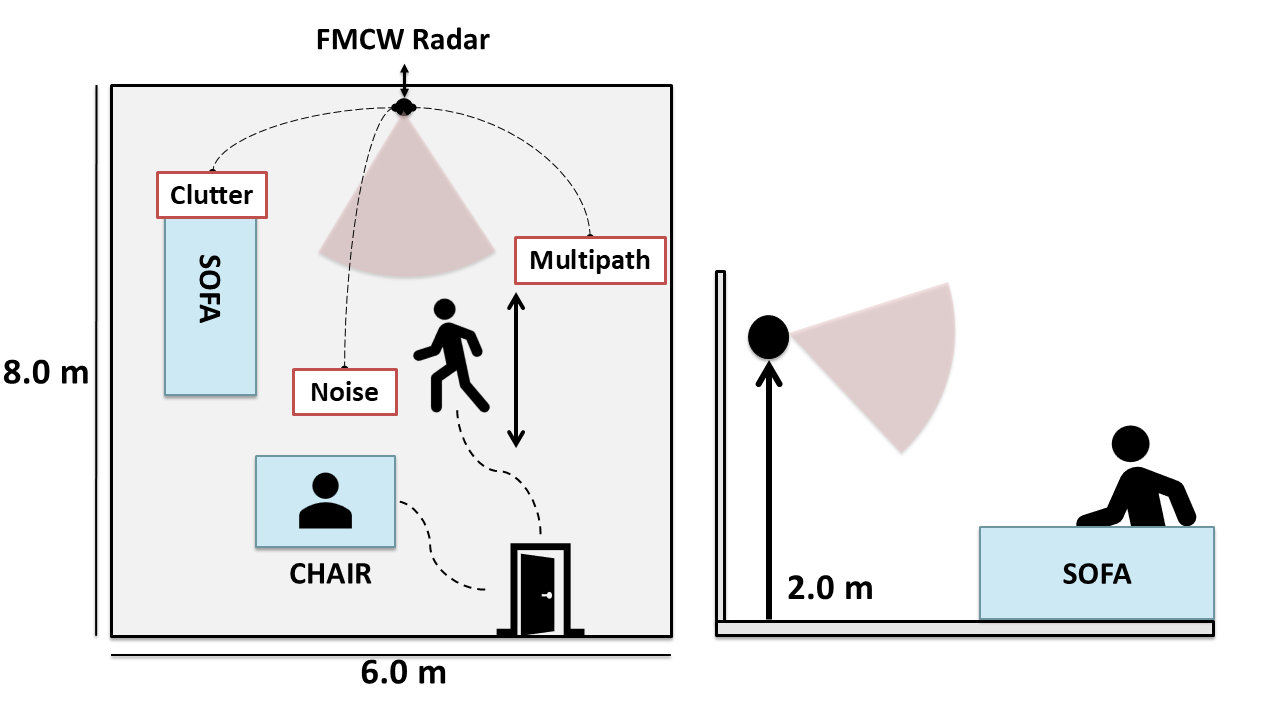}
\caption{Schematic layout of the simulated room environment (8\,m $\times$ 6\,m) including furniture (sofa, chair) and representative human activities.}
\label{fig:test_room}
\end{figure}

The corridor and room test setups are shown in Figs.~\ref{fig:test_corridor} and~\ref{fig:test_room}. Radar pose and antenna half-power beamwidths (HPBWs) match datasheet values of $30.5^\circ$ (azimuth) and $60.5^\circ$ (elevation). In the corridor, the boresight is aligned with the corridor axis ($\theta=0^\circ$), whereas in the room it is tilted by $30^\circ$ toward the activity area.

\begin{figure*}[!t]
\centering
\subfloat[]{\includegraphics[width=0.28\textwidth]{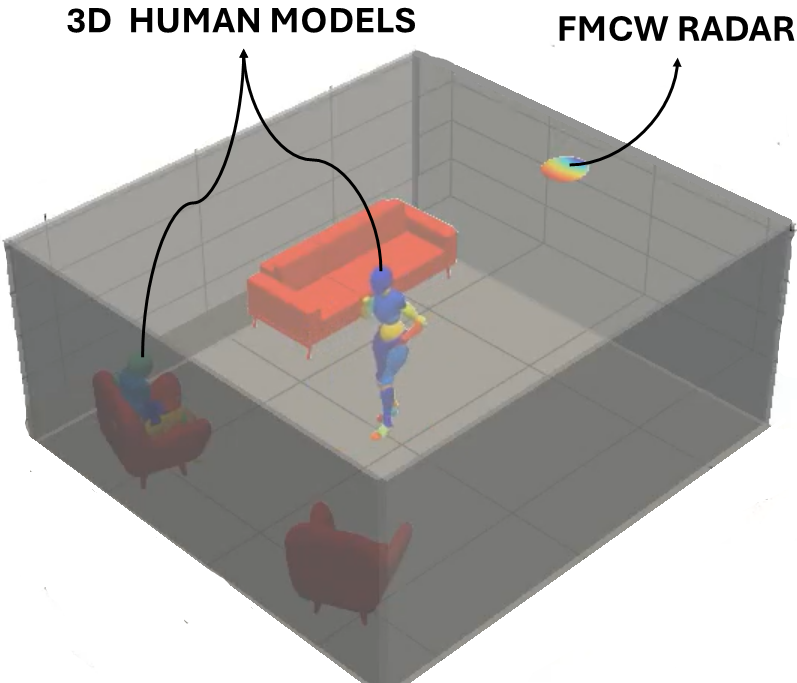}}
\hspace{0.02\textwidth}
\subfloat[]{\includegraphics[width=0.29\textwidth]{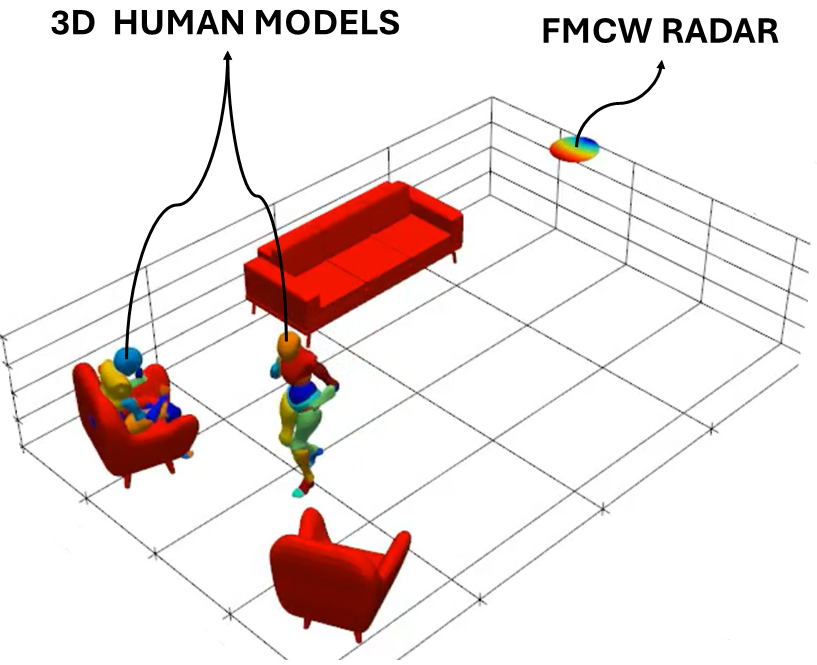}}
\hspace{0.02\textwidth}
\subfloat[]{\includegraphics[width=0.26\textwidth]{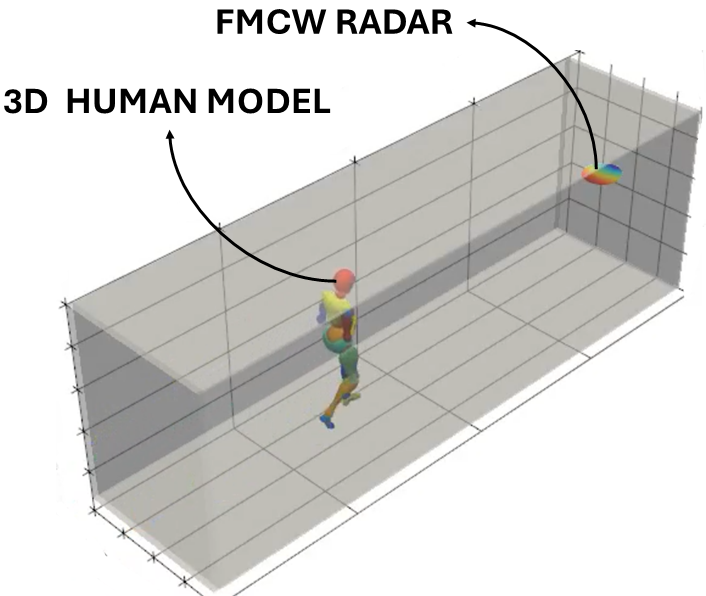}}

\vspace{3mm}

\subfloat[]{\includegraphics[width=0.28\textwidth]{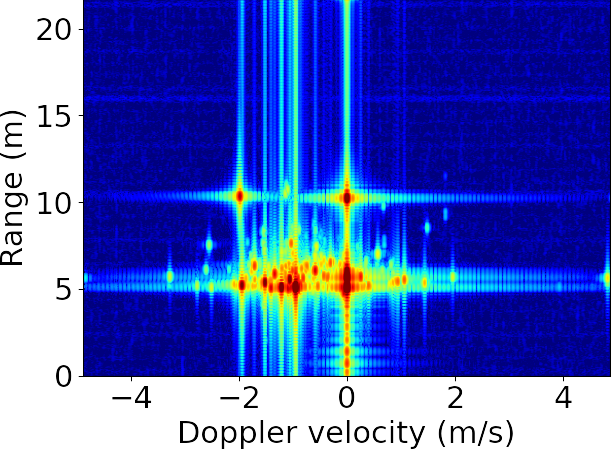}}
\hspace{0.02\textwidth}
\subfloat[]{\includegraphics[width=0.27\textwidth]{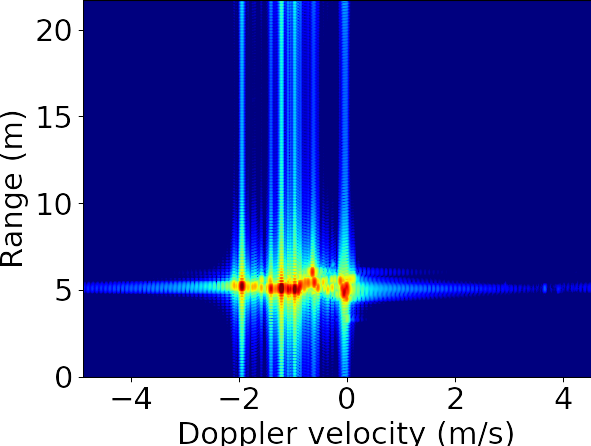}}
\hspace{0.02\textwidth}
\subfloat[]{\includegraphics[width=0.28\textwidth]{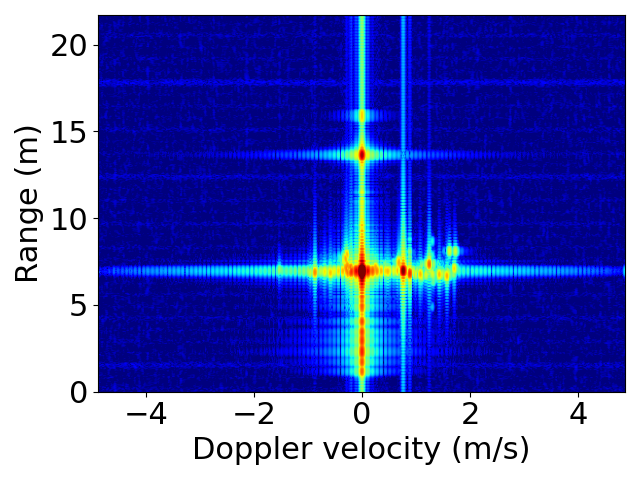}}
\caption{Simulated indoor environments and corresponding range--Doppler maps (RDMs). 
(a) Room with furniture and two occupants; (b) digital environment with furniture only; (c) corridor with one walking occupant. (d)--(f) RDMs for (a)--(c). Each RDM represents a single snapshot in time rather than the full scenario duration.}

\label{fig:dt_scene_examples}
\end{figure*}

Fig.~\ref{fig:dt_scene_examples} shows representative simulated environments alongside their corresponding RDMs, illustrating how scene geometry and human activity shape radar backscatter. The modular pipeline supports batch simulation and dataset generation at scale.

\subsection{Dataset Sizes, Splits, and Metrics}
\label{sec:data_splits_metrics}
Using the corridor and room configurations described in Sec.~\ref{sec:scenes}, the dataset comprises 2{,}650 synthetic and 1{,}000 real RDMs ($128\times128$) across three occupancy classes (empty, one person, two people). Both corpora are split $80/20$ into train/test with scene- and sequence-level separation to prevent temporal leakage. This split balances sufficient data for convergence of deeper baselines with a held-out set large enough for SNR sweeps and per-class reporting.

Within each domain (synthetic/real) and scene/occupancy cell (corridor/room), we compute the mean and standard deviation $(\mu,\sigma)$ on the training split and standardize all frames in that cell using those values (per-simulation, per-scenario standardization). This reduces scale bias while preserving the absolute-level offset reported below. To study sensitivity to additive noise in a controlled manner, we inject additive white Gaussian noise (AWGN) on held-out test frames only. For each such frame $X$, we compute its raw (pre-standardization) average power $P_x=\mathbb{E}[X^2]$, and form
\begin{equation}
Y \;=\; X + W,\qquad
W_{ij}\sim\mathcal{N}\!\left(0,\;\frac{P_x}{10^{\mathrm{SNR}_{\mathrm{dB}}/10}}\right),
\label{eq:awgn}
\end{equation}
with independent and identically distributed pixels and $\mathrm{SNR}_{\mathrm{dB}}\in\{-20,-10,+10,+20\}$. The same per-scenario standardization learned from the training split is then applied to $Y$. A fixed RNG seed ensures identical noisy realizations across models for a given file/SNR, with seeds varying across files and SNR levels.

While the SBR solver provides a high-fidelity simulation of scene-dependent effects such as clutter and multipath, the present analysis is confined to AWGN injection for controlled benchmarking. Although the simulator is capable of modeling additional hardware-specific impairments such as receiver thermal noise and oscillator phase noise, these effects are not included in the current study.
Each real scene is paired with a synthetic counterpart; training-split pixel statistics appear in Table~\ref{tab:domain_stats_compact}. Synthetic frames are approximately $170$\,dB lower than real frames, reflecting the simulator's idealized receiver chain (no absolute power or noise-floor calibration). Rather than applying a global dB offset, which could conflate calibration with classifier behavior, we (i) report performance per domain, (ii) standardize per scenario as above, and (iii) treat this offset as a key limitation motivating future work on calibrated digital twins that more accurately model the complete hardware signal chain.

\begin{table}[!htbp]
  \centering\footnotesize
  \setlength{\tabcolsep}{4pt}
  \caption{Real–synthetic pairing with per-scene pixel statistics. Values reflect dataset-level distributions, not noise injection.}
  \label{tab:domain_stats_compact}
  \begin{tabular}{l c c}
    \hline
    Scene & Real $(\mu,\sigma)$ & Synth $(\mu,\sigma)$ \\
    \hline
    Corr.—Empty      & $(-24.98,\,5.48)$ & $(-198.02,\,0.00)$ \\
    Corr.—1 person   & $(-24.55,\,5.83)$ & $(-195.00,\,4.43)$ \\
    Corr.—2 people   & $(-24.36,\,5.90)$ & $(-194.25,\,5.29)$ \\
    Room—Empty       & $(-24.63,\,5.52)$ & $(-196.26,\,0.00)$ \\
    Room—1 person    & $(-24.40,\,5.70)$ & $(-194.52,\,3.43)$ \\
    Room—2 people    & $(-24.19,\,5.67)$ & $(-194.89,\,2.72)$ \\
    \hline
  \end{tabular}
\end{table}

\paragraph*{Measurement and Simulation Campaign}
Real data was collected using an Infineon BGT60TR13C radar in two environments.  
In an institutional corridor at the University of Waterloo (length $12$\,m, width $2$\,m), the radar was mounted at $1.5$\,m height and oriented along the hallway. We recorded $40$\,s of continuous data while one and two participants entered the area, walked toward and past the radar, and repeated passes with variations in walking speed and participant height.  
In a furnished living area within a long-term care facility, the radar was mounted at $2$\,m height with a $30^\circ$ downward tilt. We recorded $50$\,s of continuous data including one person entering and sitting, another pacing in front of the radar, and each participant alone. The protocol was repeated with two independent participant groups to increase subject diversity.  
In total, this experimental campaign produced about $1{,}000$ labeled frames across the three occupancy classes.

\begin{figure}[!htbp]
  \centering
  \subfloat[]{%
    \includegraphics[width=0.45\columnwidth]{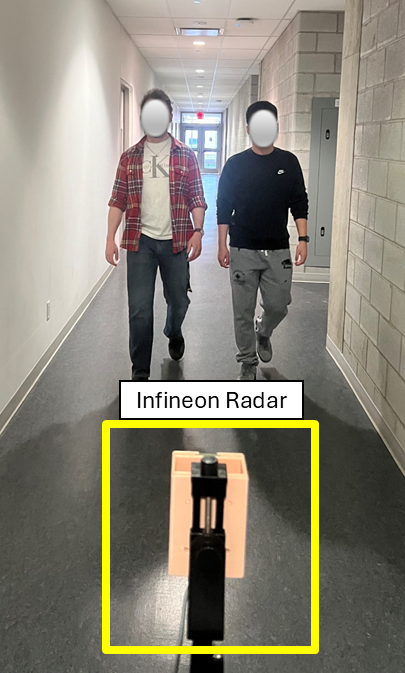}
  }
  \caption{Experimental setup for data collection in a University of Waterloo corridor, showing a 40-second walking sequence.}
  \label{fig:real_corridor}
\end{figure}

\vspace{-0.10em}
\begin{figure}[!htbp]
  \centering
  \subfloat[]{%
    \includegraphics[width=0.46\columnwidth]{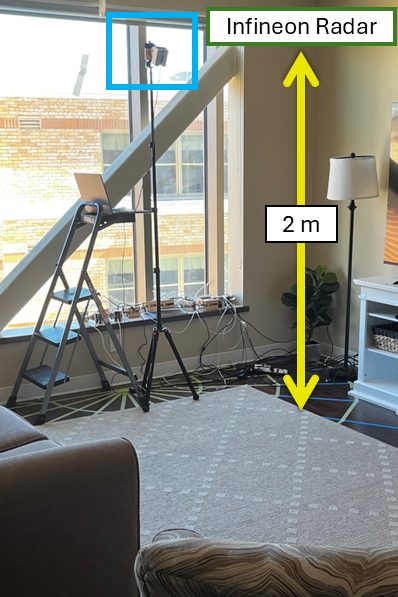}
  }\hfill
  \subfloat[]{%
    \includegraphics[width=0.46\columnwidth]{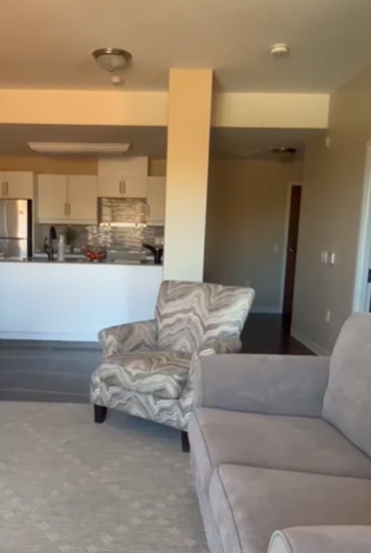}
  }
\caption{Experimental setups for real data collection in the MIRADA (``Technology for Health Empowerment: Monitoring, Intervention, and Response for Aging'') demonstration apartment at the Schlegel--UW Research Institute for Aging, operated by the Wireless Sensors and Devices Laboratory. For room measurements, the radar was mounted $2$\,m above the floor and tilted $30^{\circ}$ downward.}
\label{fig:real_room}
\end{figure}

Synthetic counterparts were generated for both environments using our simulation framework with matched corridor and room geometries and equivalent participant trajectories. To enlarge the dataset, synthetic recordings were run for slightly longer durations than their real counterparts, resulting in $2{,}650$ frames. These simulations closely mirrored the real setups but also highlight the domain gap introduced by the simulator's idealized receiver chain and lack of hardware noise.

\paragraph*{Evaluation Metrics}
Although each domain (synthetic and real) is stratified to preserve class proportions in the train/test splits, we adopt balanced accuracy (BA) as the primary metric. BA provides more informative evaluation under our noisy test conditions, where additive perturbations can disproportionately affect individual classes. In contrast, standard accuracy can mask per-class degradation and thus fails to capture the full impact of injected noise. 

Formally, BA is defined as the macro-average of per-class recall. For class $c \in \{0,1,2\}$ (empty, one person, two people), let $TP_c$ and $FN_c$ denote the number of true positives and false negatives, respectively. The per-class recall is
\begin{equation}
r_c=\frac{TP_c}{TP_c+FN_c},\quad c\in\{0,1,2\},
\end{equation}
and the balanced accuracy is
\begin{equation}
\mathrm{BA}=\tfrac{1}{3}\sum_{c=0}^{2}r_c.
\end{equation}

For completeness, we also report the macro-F1 score in Section~\ref{sec:results}, as it balances precision and recall across classes, and additionally analyze recall for the populated classes (one or two people). Denoting the union of these classes as "populated," we define
\begin{equation}
\mathrm{Rec}_{\mathrm{pop}}=\frac{TP_{\mathrm{pop}}}{TP_{\mathrm{pop}}+FN_{\mathrm{pop}}},
\end{equation}
where $TP_{\mathrm{pop}}$ and $FN_{\mathrm{pop}}$ are the true positives and false negatives aggregated over the populated classes.

\subsection{Quantum Foundations for Learning Models}

Quantum computing provides a fundamentally different framework for information processing, rooted in the postulates of quantum mechanics~\cite{Feynman1982Simulating}. At its core is the quantum bit (qubit), which generalizes the classical bit by allowing a complex linear combination of computational basis states. A qubit can thus exist in a superposition of logical zero and one,
\begin{equation}
|\psi\rangle = \alpha |0\rangle + \beta |1\rangle, \qquad |\alpha|^2 + |\beta|^2 = 1,
\end{equation}
where the absolute value squared of the complex amplitudes $\alpha$ and $\beta$ define the probabilities of measurement outcomes.

When $n$ qubits are combined, the joint state occupies a $2^n$-dimensional Hilbert space,
\begin{equation}
|\Psi\rangle = \sum_{j=0}^{2^n - 1} \alpha_j |j\rangle,
\end{equation}
where $|j\rangle$ enumerates all binary configurations of $n$ qubits. This exponential state space enables compact representation of high-dimensional relationships that are intractable for classical models.

Two essential quantum resources for learning are entanglement and interference. Entanglement generates non-classical correlations between qubits, enabling encoding of non-separable patterns. A canonical example is the Bell state,
\begin{equation}
|\Phi^+\rangle = \tfrac{1}{\sqrt{2}} \big( |00\rangle + |11\rangle \big),
\end{equation}
which exhibits perfect correlation between qubits. Interference governs how probability amplitudes combine; by adjusting relative phases, certain outcomes are amplified while others are suppressed. Together, these effects support richer hypothesis spaces for modeling dependencies in data.

To extract information from a quantum system, measurements are performed. Direct measurement collapses the state to a single basis outcome, and in learning applications, expectation values of Hermitian operators (observables) are typically computed:
\begin{equation}
\langle O \rangle = \langle \Psi | O | \Psi \rangle,
\end{equation}
which provide statistical summaries of the state and can be used as features for downstream classification. As they represent statistical averages, expectation values are estimated by repeated measurement of identically prepared copies of a given quantum state.

Parameterized quantum circuits (PQCs) form the backbone of quantum machine learning. PQCs apply sequences of unitary gates with tunable parameters, optimized to minimize a classical loss function, making them quantum analogues of neural networks. A critical design choice is feature mapping: embedding classical data into quantum states. Approaches range from angle encoding to entangling feature maps such as the ZZFeatureMap~\cite{Havlicek2019Supervised}, which incorporates pairwise feature interactions via controlled rotations. The chosen embedding defines the geometry of the subset of Hilbert space explored during training and directly impacts representational power.
While the expressive power of PQCs has motivated research into formal quantum advantage, the emphasis in this work is on practical quantum utility. Specifically, we investigate whether a compact, quantum-informed model can provide tangible benefits, such as parameter efficiency and improved stability under additive noise, for a targeted sensing task, even when it does not exceed the peak performance of established classical baselines. Given the limited qubit counts, coherence times, and error rates of current Noisy Intermediate-Scale Quantum (NISQ) devices, we conduct all experiments on a classical backend which lets us setup the foudnations for future hardware tests. Specifically, circuits are simulated with Qiskit Aer in statevector (analytic, noise-free) mode~\cite{Qiskit2024}, enabling controlled evaluation of model capacity separate from hardware noise.
These foundations motivate the minimalist two-qubit hybrid in Sec.~\ref{sec:qnn_arch}: a 2-D latent is embedded with a ZZFeatureMap, transformed by a depth-1 RealAmplitudes ansatz, and read out via expectation values of local Pauli observables (\(ZI, IZ\)) to form a two-dimensional quantum feature for classification.

\subsection{Quantum Neural Network Architecture}
\label{sec:qnn_arch}
Quantum neural networks (QNNs) extend classical models by inserting a PQC between feature extraction and classification. A typical QNN comprises three stages, data embedding, variational transformation, and measurement.

\begin{figure}[t]
\centering
\includegraphics[width=0.93\columnwidth]{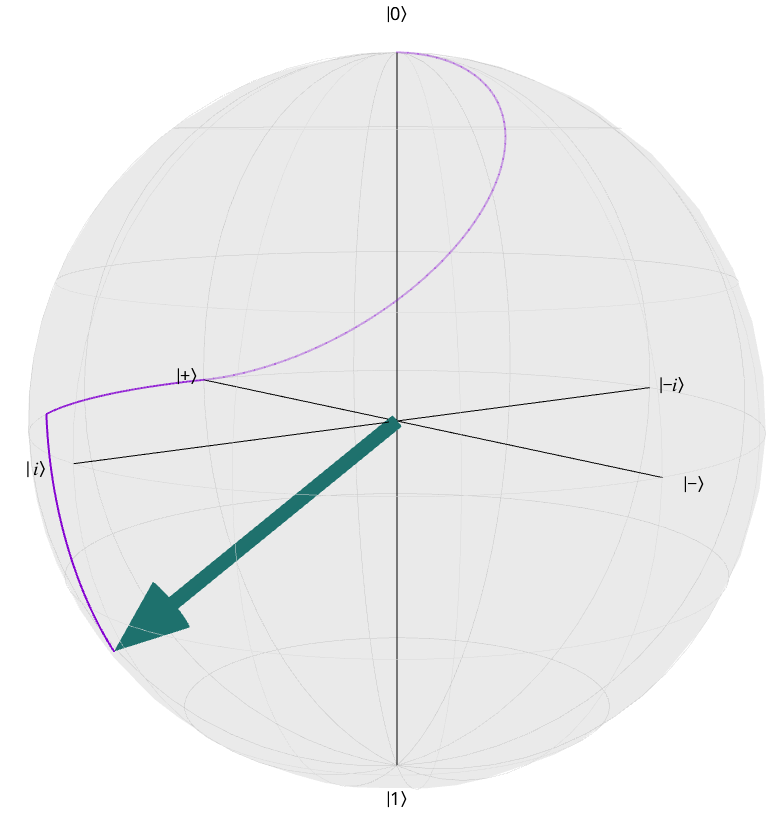}
\caption{Bloch-sphere visualization of the $[0,\pi]$ input angle normalization. 
Starting from $\ket{0}$ at the north pole, successive $R_y(\theta)$ rotations sweep the state monotonically toward $\ket{1}$ along a tilted meridian (illustrated here with the toy sequence $H \rightarrow P(\phi) \rightarrow R_y(\theta)$ for $\phi=\pi/4$). 
This encoding covers the full pole-to-pole range without redundancy, whereas symmetric choices such as $[-\pi/2,\pi/2]$ remain confined to the equator. 
Figure generated using the open-source Bloch Sphere visualization tool~\cite{blochsphereapp}; the sequence reflects the embedding and rotation structure of the full HQNN PQC shown in Fig.~\ref{fig:pqc_diagram}.}
\label{fig:bloch_meridians}
\end{figure}

In this work, each $128\times128$ RDM is processed by a compact CNN frontend to produce a latent $\mathbf{x}=(x_0,x_1)\in\mathbb{R}^2$. Prior to angle mapping, each latent component is standardized using the training-set statistics, $x_i \leftarrow (x_i-\mu_i)/\sigma_i$, to stabilize optimization and avoid outlier-driven angles. These values are then squashed into $[0,\pi]^2$ via the logistic function,
\begin{equation}
\boldsymbol{\theta}_{\mathrm{in}}=\pi\,\sigma(\mathbf{x})\in[0,\pi]^2,
\end{equation}
so that each qubit is placed on a Bloch-sphere meridian sweeping monotonically from the north pole ($|0\rangle$) to the south pole ($|1\rangle$). 
We restrict the domain to $[0,\pi]$ rather than $[0,2\pi]$ to avoid redundant 
wraparound, since $0$ and $2\pi$ correspond to the same physical state. The 
interval $[-\pi/2,\pi/2]$ is also avoided, as it confines the state to a single hemisphere and never reaches $\ket{1}$. This choice ensures that each latent dimension produces a unique pole-to-pole trajectory, yielding encodings that are both geometrically interpretable and empirically more stable during training. A bounded, monotone logistic map prevents extreme inputs from producing unstable rotations while still covering the full pole-to-pole range. Fig.~\ref{fig:bloch_meridians} illustrates the effect of these normalizations, with trajectories generated by a toy sequence.

We implement a compact two-qubit quantum circuit that alternates between data encoding and trainable processing. First, the input angles $\boldsymbol{\theta}_{\mathrm{in}}$ are embedded into the qubits using a feature-mapping block (commonly implemented as a ZZFeatureMap in Qiskit). This step rotates the qubits according to the input values, effectively imprinting the data onto the quantum system. The encoded state is then processed by a shallow variational layer (a RealAmplitudes ansatz with one repetition and full entanglement in Qiskit), which introduces four trainable rotation parameters. Together, these steps prepare the quantum state
\begin{equation}
|\psi(\mathbf{x};\boldsymbol{\theta})\rangle
= U(\boldsymbol{\theta})\,\Phi(\boldsymbol{\theta}_{\mathrm{in}})\,|00\rangle .
\label{eq:qnn_state}
\end{equation}

From this state, we compute expectation values of Pauli-$Z$ operators on each qubit. Here, $ZI$ denotes applying $Z$ to the first qubit and identity $I$ to the second, while $IZ$ denotes the reverse. These measurements yield two real numbers in $[-1,1]$, which together form a two-dimensional quantum feature vector
\begin{equation}
\mathbf{f}(\mathbf{x};\boldsymbol{\theta})=
\begin{bmatrix}
\langle \psi \vert ZI \vert \psi \rangle\\[2pt]
\langle \psi \vert IZ \vert \psi \rangle
\end{bmatrix}\in[-1,1]^2 .
\label{eq:qnn_obs}
\end{equation}

This feature vector is passed to a classical linear layer with trainable weights $W \in \mathbb{R}^{C\times 2}$ and bias $\mathbf{b}\in\mathbb{R}^C$, producing a vector of logits
\begin{equation}
\mathbf{z}=W\mathbf{f}+\mathbf{b},
\end{equation}
where $z_c$ denotes the logit associated with class $c$. A softmax converts these logits into class probabilities
\begin{equation}
p_c=\frac{e^{z_c}}{\sum_{c'=1}^{C}e^{z_{c'}}},\qquad c=1,\dots,C ,
\end{equation}
with $C{=}3$ classes in our case.

Both the quantum parameters $\boldsymbol{\theta}$ and the classical parameters $(W,\mathbf{b})$ are optimized jointly during training. We use a class-weighted cross-entropy loss,
\begin{equation}
\mathcal{L}=-\sum_{c=1}^{C}w_c\,y_c\log p_c \, ,
\end{equation}
where $\mathbf{y}\in\{0,1\}^C$ is the one-hot ground-truth vector, $y_c$ is its $c$-th component, $p_c$ is the predicted probability for class $c$, and $w_c$ is the weight assigned to class $c$.

The quantum component of our model is a deliberately shallow PQC with two qubits and a single repetition (\texttt{reps}=1), containing only four trainable rotation angles (Fig.~\ref{fig:pqc_diagram}). This minimal design was chosen to control runtime and avoid over-expressivity. For circuits of this scale, classical simulation on a CPU proved more efficient than on a GPU, and CPUs were therefore used for all final experiments.

Expectation values were evaluated using the Qiskit-Aer Estimator in statevector simulation mode. Gradients for the quantum parameters were obtained via the analytic parameter-shift rule, while the classical CNN branch was trained using standard backpropagation~\cite{Schuld2019}. Unlike real quantum hardware, which must estimate expectation values from repeated measurements ("shots"), the statevector simulator can also compute them exactly and without noise when configured with \texttt{shots=None}. By default, however, the Estimator mimics hardware by sampling 4096 shots, which introduces artificial statistical fluctuations that scale as $1/\sqrt{\text{shots}}$. In our experiments, we used this default shot-based setting (4096 shots), so our reported values include these hardware-like fluctuations~\cite{Qiskit2024}.

\subsection{Model Architectures and Training}
\label{sec:models_training}
We evaluate four models spanning task-specific CNNs, general-purpose deep architectures, and a hybrid quantum--classical neural network. The baseline is a DopplerNet-style CNN tailored for micro-Doppler radar classification, consisting of five convolutional blocks and three fully connected layers, adapted to adapted to $128\times128$ RDMs~\cite{Zhang2020DopplerNet}, and serves as a radar-specific baseline classification tasks.

To benchmark against deeper classical architectures, we include ResNet-18~\cite{He2016ResNet} and EfficientNet-B0~\cite{Tan2019EfficientNet} with pre trained weights. ResNet-18 employs skip connections and has shown strong performance in FMCW radar tasks such as human activity recognition~\cite{Wang2021RadarHAR}. EfficientNet-B0 uses compound scaling to balance depth, width, and resolution. Both are adapted to single-channel radar inputs and initialized with ImageNet weights; EfficientNet-B0 additionally uses a $3\times3\times3$ stem convolution for early spatial aggregation. Together with DopplerNet, these represent lightweight and high-capacity CNN designs. Together with DopplerNet, these models span both radar-specialized and general-purpose CNN designs that have been widely adopted in FMCW radar applications.

\begin{figure*}[!t]
  \centering
  \includegraphics[width=1.5\columnwidth]{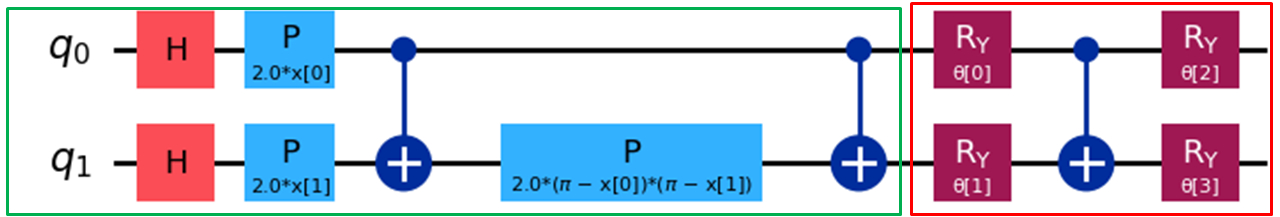}
  \caption{Parameterized quantum circuit used in HQNN. The left block implements the ZZFeatureMap (data embedding and entanglement); the right block applies the RealAmplitudes ansatz with trainable rotations and entangling gates.}
  \label{fig:pqc_diagram}
\end{figure*}

\begin{figure*}[!t]
  \centering
  \includegraphics[width=1.5\columnwidth]{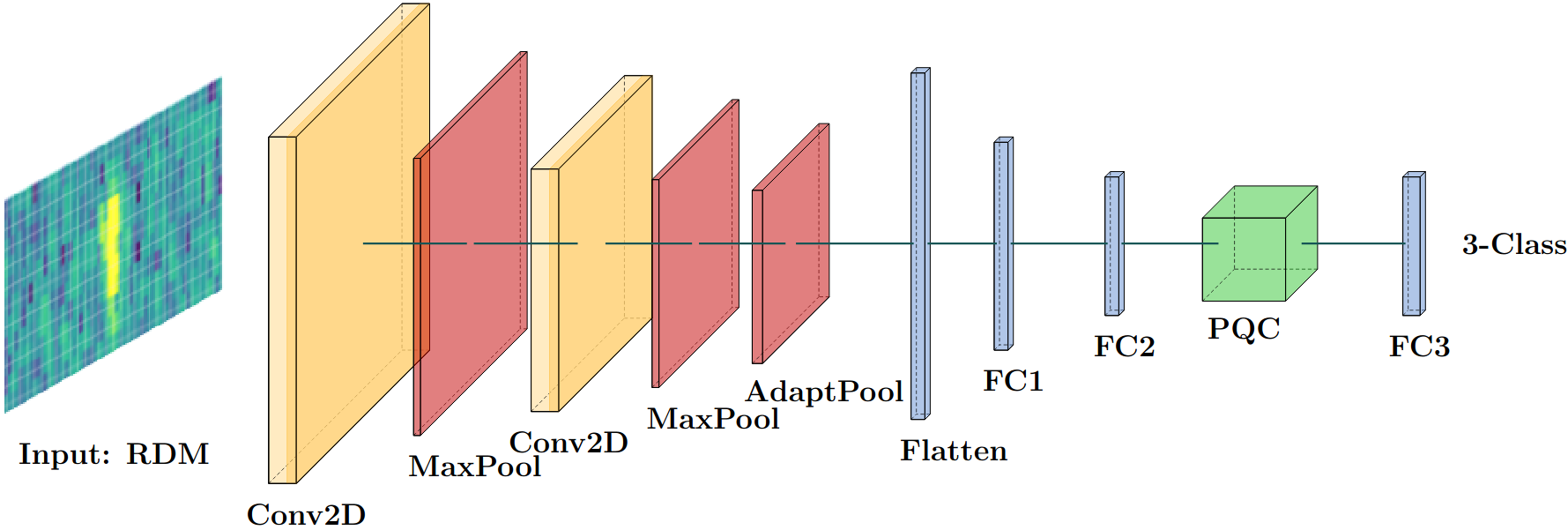}
  \caption{Architecture of the Hybrid Quantum--Classical Neural Network.}
  \label{fig:hqnn_architecture}
\end{figure*}

For hybrid learning, we employ the 2-qubit HQNN of Sec.~\ref{sec:qnn_arch}. A compact CNN frontend extracts a two-dimensional latent vector, which is embedded into a quantum state by a ZZFeatureMap and transformed by a RealAmplitudes ansatz (reps$=1$), full entanglement; four trainable rotation angles). Expectation values $\{ZI,\,IZ\}$ per~\eqref{eq:qnn_obs} form the quantum feature vector, which a linear layer maps to three logits. Notably, this architecture intentionally avoids classical complexity such as the skip connections and deep stage hierarchy of ResNet-18 or the compound scaling of EfficientNet, delegating nonlinear feature mixing to the PQC; this keeps the classical path shallow and the parameter budget minimal while isolating the contribution of the quantum layer within our noise-sensitivity protocol. Fig.~\ref{fig:pqc_diagram} shows the PQC; Fig.~\ref{fig:hqnn_architecture} illustrates the full HQNN pipeline. 

\begin{table}[!t]
  \caption{HQNN layer-by-layer specification (shared CNN backbone, 2-qubit PQC).
  Trainable parameters include biases; non-trainable steps are shown with 0.}
  \label{tab:hqnn_spec}
  \centering
  \footnotesize
  \setlength{\tabcolsep}{3pt} 
  \renewcommand{\arraystretch}{1.05}
  \begin{tabularx}{\columnwidth}{@{}l X c r@{}}
    \hline
    Stage & Configuration & Output & Params \\
    \hline
    Input & $1{\times}128{\times}128$ RDM & $1{\times}128{\times}128$ & -- \\
    Conv2D $\to$ MaxPool & $k{=}3$, $p{=}1$, $1{\to}16$; pool $=2$ & $16{\times}64{\times}64$ & 160 \\
    Conv2D $\to$ MaxPool & $k{=}3$, $p{=}1$, $16{\to}32$; pool $=2$ & $32{\times}32{\times}32$ & 4,640 \\
    Adaptive AvgPool & target $(2,15)$ & $32{\times}2{\times}15$ & 0 \\
    Flatten & -- & 960 & 0 \\
    FC1 & $960{\to}64$ & 64 & 61,504 \\
    FC2 & $64{\to}2$ & 2 & 130 \\
    Angle squash & $\sigma(\cdot)\cdot\pi$ & 2 & 0 \\
    PQC (Estimator) &
      ZZFeatureMap $+$ RealAmplitudes ($\texttt{reps}=1$) & 2 & 4 \\
    FC3 (head) & $2{\to}3$ & 3 & 9 \\
    \hline
    \multicolumn{3}{r}{\textbf{Total trainable}} & \textbf{66,447} \\
    \hline
  \end{tabularx}
\end{table}

We considered QSVMs~\cite{schuld2019quantumML}, amplitude-encoding classifiers, and variational algorithms such as VQE/QAOA~\cite{Biamonte2017QuantumML}, but excluded them for pragmatic and methodological reasons. Amplitude encoding would require preparing a $d{=}16384$-dimensional state for each $128{\times}128$ frame and, in general, incurs substantial data-loading/state-preparation overhead that can dominate runtime in practice~\cite{Biamonte2017QuantumML}. Trainability concerns further motivate a shallow two-qubit variational block: gradients in deeper/wider circuits degrade rapidly in the presence of noise and random initializations~\cite{wang2021noise}. Fully quantum classifiers also remain impractical at current qubit counts and coherence times~\cite{Beer2020QNN}. In contrast, the proposed HQNN preserves radar structure via a compact CNN, keeps the quantum component small (four trainable parameters), enables end-to-end gradients, and is compatible with near-term devices.

\subsubsection{Training setup and runtime}
\label{sec:training_runtime}
All models are trained under identical optimization settings in both synthetic and real domains: Adam optimizer (learning rate $10^{-3}$), batch size $32$, and class-weighted cross-entropy loss. Classical CNNs are implemented in PyTorch, while the HQNN leverages Qiskit's TorchConnector~\cite{QiskitMachineLearning} with the Aer Estimator~\cite{Qiskit2024}. Quantum simulation is performed in noiseless statevector mode with Qiskit Aer and optional NVIDIA cuStateVec acceleration (RTX~4090 GPU, Intel i9-12900K CPU, 32~GB RAM). Training is run for 15 epochs on synthetic data and 25 epochs on real data to ensure convergence, with identical AWGN configurations applied to the held-out test sets for zero-shot robustness evaluation.

Table~\ref{tab:setup_params} summarizes parameter budgets and training times for the main models. 
The HQNN contains roughly 66 thousand parameters, which is around 40 times fewer than DopplerNet and $170\times$ less than ResNet-18, yet it incurs substantially longer training times due to PQC evaluation and classical--quantum tensor handoffs.

\begin{table}[!h]
  \caption{Parameter counts and wall-clock training times for the main models (Synthetic: 15 epochs, Real: 25 epochs).}
  \label{tab:setup_params}
  \centering\footnotesize
  \begin{tabular}{l r r r}
    \hline
    Model           & Params (M) & Synthetic (min) & Real (min) \\
    \hline
    HQNN            & 0.066      & 11.85           & 6.95 \\
    EfficientNet-B0 & 4.011      & 0.26            & 0.41 \\
    ResNet-18       & 11.172     & 0.25            & 0.31 \\
    DopplerNet      & 2.558      & 0.25            & 0.29 \\
    \hline
  \end{tabular}
\end{table}

Profiling indicates that the quantum forward pass dominates batch latency, at approximately $11$–$16$\,ms per batch with the Estimator, compared to about $0.7$\,ms per batch for a classical head. An additional measurable overhead arises from CPU--GPU transfers of the angle tensor. Slow forward pass evolution is a symptom of simulating quantum circuits on a classical computer, and will be alleviated when running on real quantum computers, especially for larger width circuits. Reducing the data transfer time between the quantum processing unit (QPU) and the CPU/GPU is a more insidious issue that plagues any heterogeneous computing framework, but is the focus of much focused research and development. 

We therefore report performance in terms of parameter efficiency and stability under additive noise, not as any claim of quantum speedup.
To separate quantum-layer contribution from classical bottlenecks, we define two lightweight control variants that retain the CNN backbone and 2-D bottleneck but replace the PQC with classical heads of comparable size. The first, uses a direct linear mapping from the 2-D bottleneck to class logits (Estimator-matched). 
The second, employs a small multilayer perceptron with one hidden layer of size $h{=}2$, chosen so that the total parameter count is comparable to the PQC case. These controls help attribute any observed differences to the PQC rather than the bottleneck itself. 
Head-only parameter counts are listed in Table~\ref{tab:ctrl_heads}, while their performance is reported in Sec.~\ref{sec:results}.

\begin{table}[!t]
  \caption{Classical control heads (shared CNN backbone and 2-D bottleneck). Parameter counts are for the head only.}
  \label{tab:ctrl_heads}
  \centering
  \footnotesize
  \setlength{\tabcolsep}{3pt}
  \renewcommand{\arraystretch}{1.05}
  \begin{tabularx}{\columnwidth}{@{}l X r@{}}
    \hline
    Head & Mapping & Params \\
    \hline
    Estimator-matched &
      $\mathrm{Linear}(2\!\to\!3)$ & 9 \\
    Dequantization (analog) &
      \begin{tabular}[t]{@{}l@{}}
        $\mathrm{Linear}(2\!\to\!h)\,{+}\,\mathrm{ReLU}$\\
        $\mathrm{Linear}(h\!\to\!3)$
      \end{tabular} & 15 \\
    \hline
  \end{tabularx}

  \vspace{2pt}
\end{table}

In practice, both control heads trained quickly (under 0.3 minutes for both 15-epoch synthetic and 25-epoch real runs), comparable to the CNN baselines. Yet despite sharing the same CNN backbone and having nearly identical parameter counts (Table~\ref{tab:ctrl_heads}), their accuracy collapses entirely, as detailed in Sec.~\ref{sec:results}. This stark failure decisively demonstrates that the PQC contributes essential representational capacity beyond what parameter-matched classical heads can provide~\cite{abbas2021power,schuld2021train, PhysRevResearch.3.013077}.

\section{Results}
\label{sec:results}

\subsection[Synthetic-Data Training Results (representative run ★)]%
{Synthetic-Data Training Results (representative run \texorpdfstring{$\bigstar$}{*})\footnotemark}
\label{subsec:synth-results}

Models were trained on the digital-twin dataset without noise injection for 15 epochs. 
In a representative held-out split, all CNN baselines and the HQNN reach above $99\%$ accuracy (Table~\ref{tab:synth_clean}), confirming that both families are sufficiently expressive for the synthetic domain. In contrast, the attribution controls (Estimator-matched and Dequantization) obtain similar accuracy on train and test splits but remain far below the performance of the CNNs and HQNN. This pattern indicates underfitting rather than overfitting, suggesting that these shallow classical heads cannot capture the task structure as effectively. While we cannot definitively ascribe the gap to model capacity alone, the results are consistent with the PQC contributing useful nonlinear transformations beyond the 2-D bottleneck.

\begin{figure*}[!t]
  \centering
  \subfloat[One person]{%
    \includegraphics[width=0.9\textwidth]{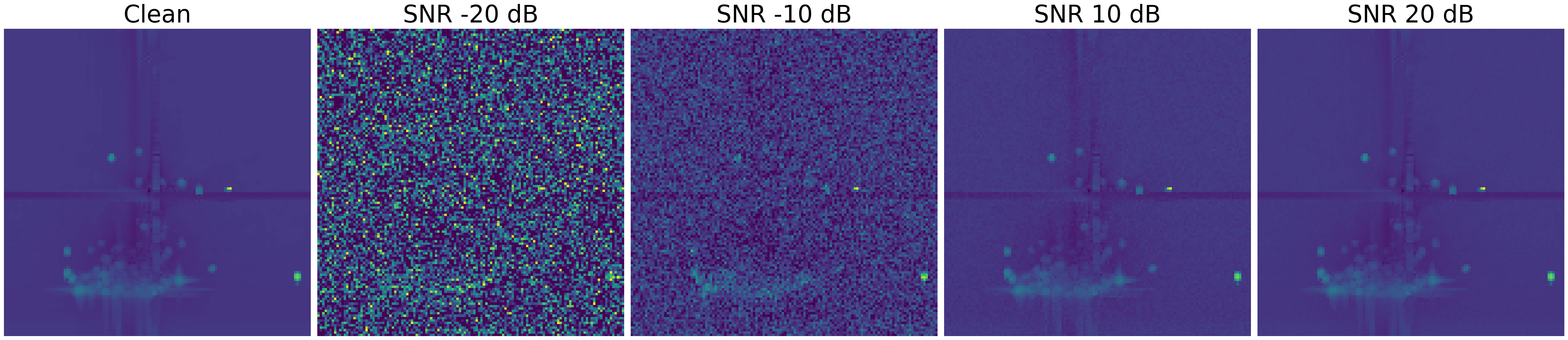}
  }\\[6pt]
  \subfloat[Two people]{%
    \includegraphics[width=0.9\textwidth]{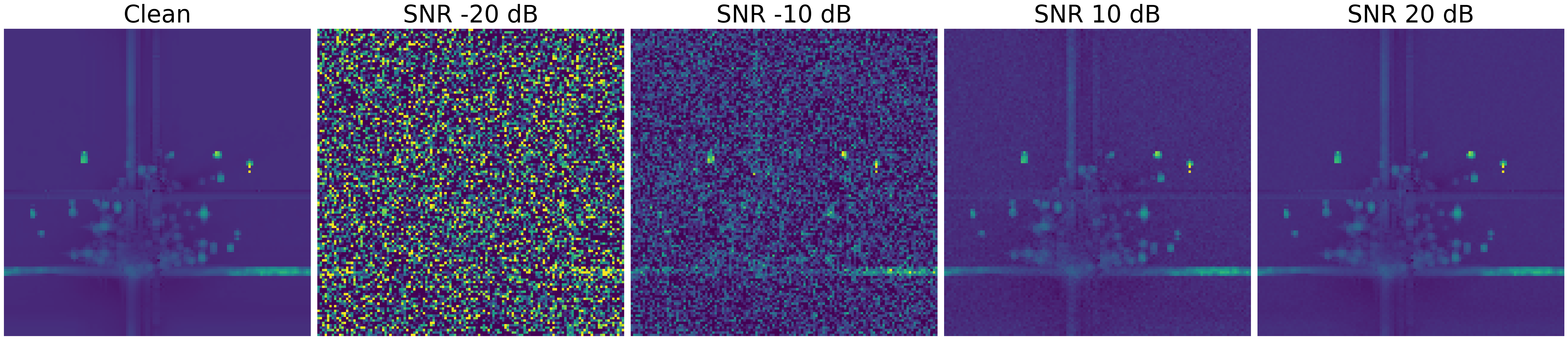}
  }\\[6pt]
  \subfloat[Empty]{%
    \includegraphics[width=0.9\textwidth]{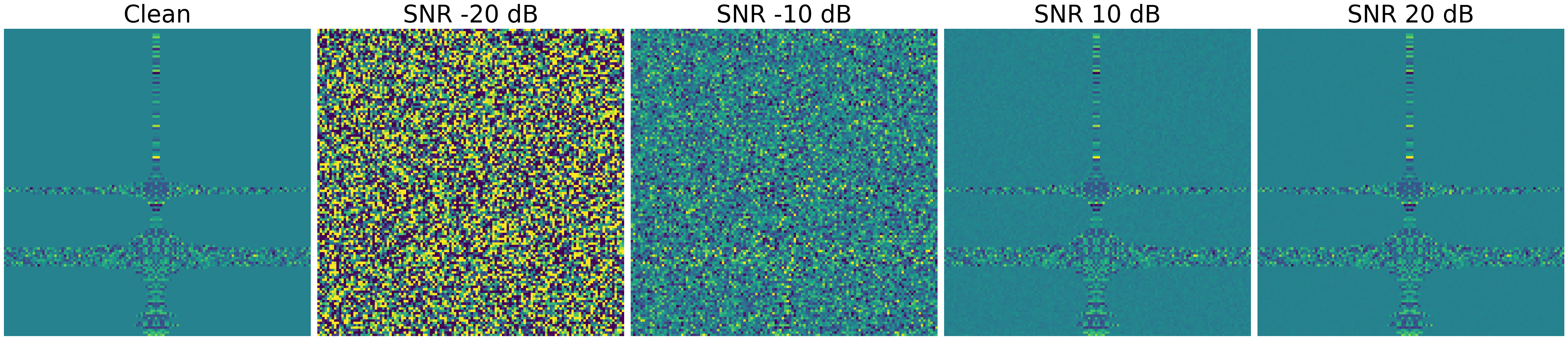}
  }
\caption{Synthetic data (representative run): range--Doppler maps under AWGN injection at different SNRs.}
  \label{fig:snr_visuals_synth}
\end{figure*}

\begin{table}[!t]
  \caption{Synthetic data: clean-set accuracy on the held-out split (representative run).}
  \label{tab:synth_clean}
  \centering\footnotesize
  \begin{tabular}{lcc}
    \hline
    Model                   & Test Accuracy & Train Accuracy \\
    \hline
    DopplerNet              & 99.9\% & 99.8\% \\
    ResNet-18               & 99.8\% & 100.0\% \\
    EfficientNet-B0         & 99.6\% & 99.9\% \\
    HQNN                    & 99.7\% & 100\% \\
    Estimator-matched       & 71.7\% & 73.8\% \\
    Dequantization (analog) & 60.2\% & 55.4\% \\
    \hline
  \end{tabular}
\end{table}
\paragraph*{Noise sensitivity using the synthetic dataset}
We next examine behavior under AWGN, injected at test time per~\eqref{eq:awgn} across SNRs $\{-20,-10,+10,+20\}$\,dB. 
Performance is summarized via Accuracy, Balanced Accuracy (BA), and Macro-F1. Representative synthetic RDMs are shown in Fig.~\ref{fig:snr_visuals_synth}: at $-20$ and $-10$\,dB reflections collapse into speckle, while at positive SNRs Doppler streaks re-emerge.  

\begin{figure*}[!t]
  \centering
  \includegraphics[width=0.98\textwidth]{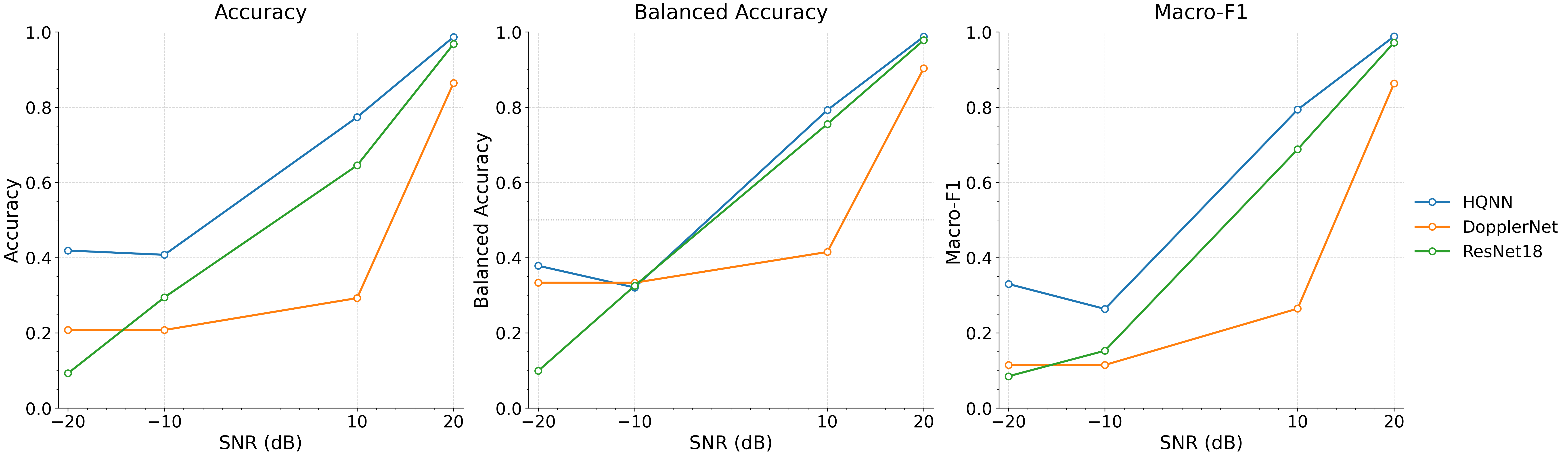}
  \caption{Synthetic data (representative run): test performance with injected AWGN, showing Accuracy, Balanced Accuracy, and Macro-F1 vs.\ SNR.}
  \label{fig:metrics_vs_snr_synth}
\end{figure*}

The overall performance trends are summarized in Fig.~\ref{fig:metrics_vs_snr_synth} and Table~\ref{tab:synth_results}. While all models perform near chance level at $-20$\,dB, the HQNN begins its recovery earlier in the mid-SNR regime ($+10$\,dB). At the highest SNR of $+20$\,dB, HQNN and ResNet-18 both achieve strong performance, while the other baselines remain lower.  

\begin{figure*}[!t]
  \centering
  \subfloat[]{%
    \includegraphics[width=0.7\textwidth]{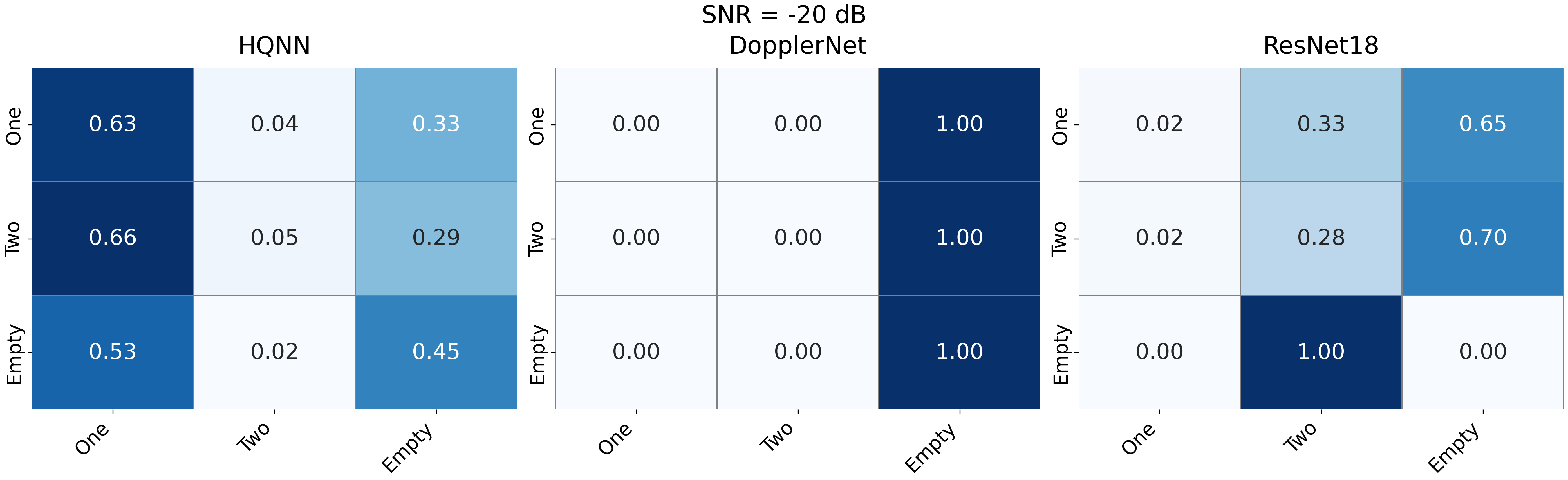}
  }\\[10pt]
  \subfloat[]{%
    \includegraphics[width=0.7\textwidth]{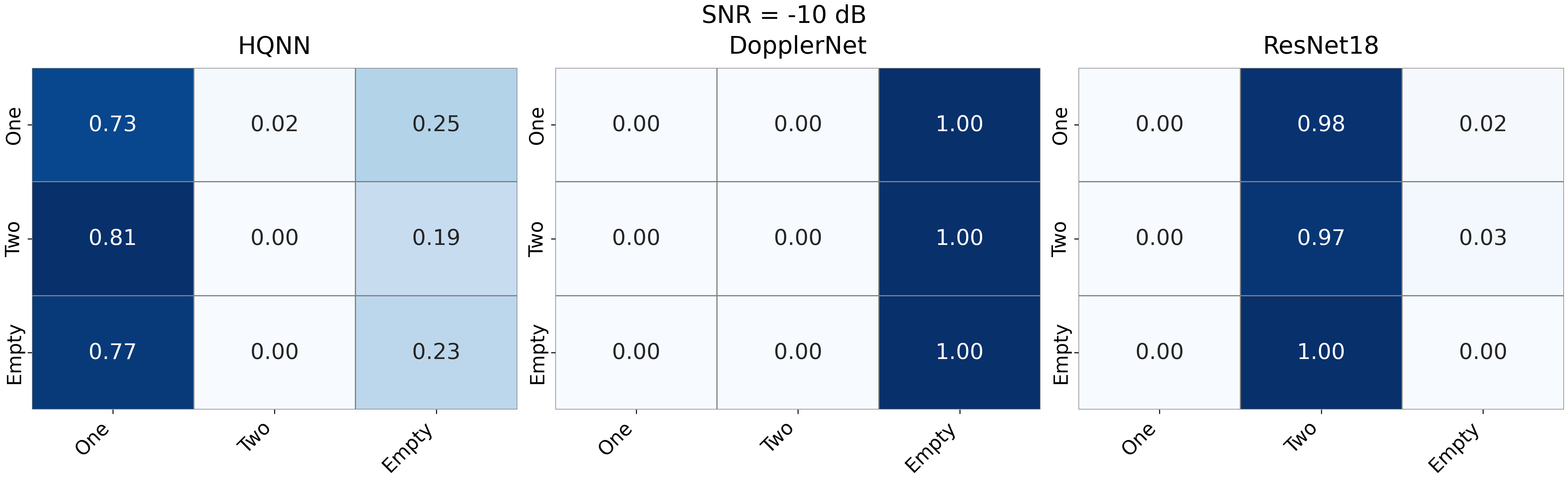}
  }\\[10pt]
  \subfloat[]{%
    \includegraphics[width=0.7\textwidth]{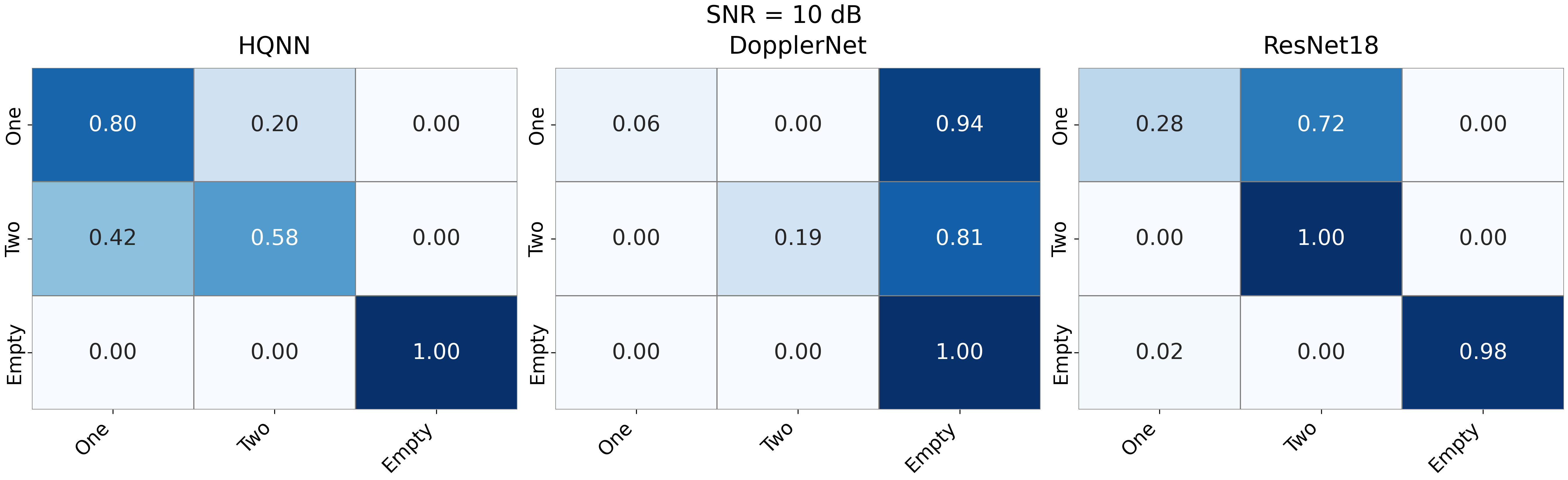}
  }\\[10pt]
  \subfloat[]{%
    \includegraphics[width=0.7\textwidth]{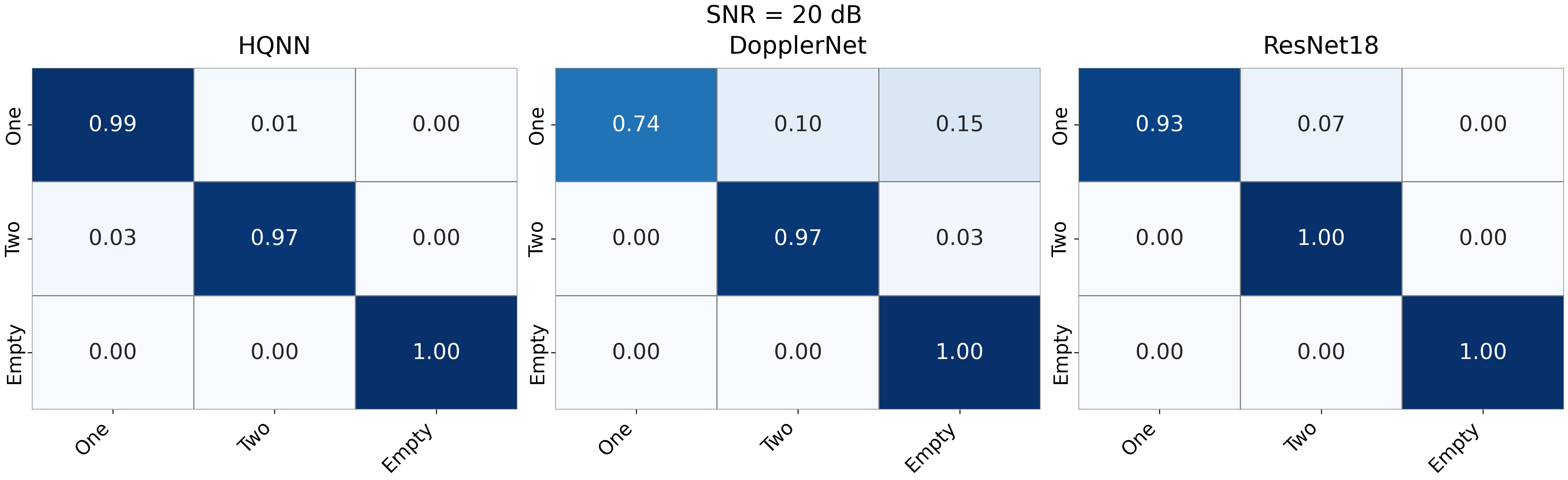}
  }
\caption{Synthetic data (representative run): normalized confusion matrices for HQNN, DopplerNet, and ResNet-18 under different SNRs.}
  \label{fig:cm_synth}
\end{figure*}

Confusion matrices for HQNN, DopplerNet, and ResNet-18 are shown in Fig.~\ref{fig:cm_synth}. At the lowest SNR of $-20$\,dB, all models degrade to chance-level predictions, with DopplerNet strongly biased toward the empty class. At $-10$\,dB, HQNN retains partial separation of one- and two-person scenes, whereas CNNs collapse (DopplerNet to "two," ResNet-18 to "one"). At $+10$\,dB, HQNN achieves $\mathrm{BA}\!\approx\!0.792$ and $\mathrm{F1}\!\approx\!0.794$, modestly ahead of ResNet-18 ($\mathrm{BA}\!\approx\!0.755$) and substantially above EfficientNet-B0 ($\mathrm{BA}\!\approx\!0.594$) and DopplerNet ($\mathrm{BA}\!\approx\!0.415$). At $+20$\,dB, all models recover strongly: HQNN and ResNet-18 both exceed $96\%$ accuracy, while EfficientNet-B0 and DopplerNet remain lower ($\approx 86$–90\%).  

\footnotetext{The star (★) denotes the highest-performing seed run, highlighted in Fig.~\ref{fig:metrics_vs_snr_synth} and Fig.~\ref{fig:metrics_vs_snr_real}. This representative run is shown for qualitative illustration; multi-seed averages with confidence intervals are reported separately in this section.}

\begin{table*}[!t]
  \caption{Synthetic data (representative run): Balanced Accuracy (BA), Macro-F1, and Accuracy across SNRs with injected AWGN.}
  \label{tab:synth_results}
  \centering
  \footnotesize
  \setlength{\tabcolsep}{6pt}
  \renewcommand{\arraystretch}{1.2}
  \begin{tabular}{l|ccc|ccc|ccc|ccc}
    \hline
    \multirow{2}{*}{Model}
      & \multicolumn{3}{c|}{$-20$ dB}
      & \multicolumn{3}{c|}{$-10$ dB}
      & \multicolumn{3}{c|}{$+10$ dB}
      & \multicolumn{3}{c}{$+20$ dB} \\
    \cline{2-13}
      & BA & F1 & Acc & BA & F1 & Acc & BA & F1 & Acc & BA & F1 & Acc \\
    \hline
    DopplerNet
      & 0.333 & 0.115 & 0.208
      & 0.333 & 0.115 & 0.208
      & 0.415 & 0.265 & 0.292
      & 0.904 & 0.863 & 0.864 \\
    ResNet-18
      & 0.333 & 0.160 & 0.313
      & 0.325 & 0.152 & 0.294
      & 0.755 & 0.688 & 0.645
      & 0.978 & 0.972 & 0.968 \\
    EfficientNet-B0
      & 0.407 & 0.389 & 0.481
      & 0.280 & 0.223 & 0.294
      & 0.594 & 0.464 & 0.475
      & 0.903 & 0.873 & 0.860 \\
    HQNN
      & 0.378 & 0.330 & 0.419
      & 0.321 & 0.264 & 0.408
      & 0.792 & 0.794 & 0.774
      & 0.988 & 0.988 & 0.987 \\
    \hline
  \end{tabular}
\end{table*}

\subsubsection{Multi-seed analysis}
\begin{figure}[!t]
  \centering
  \includegraphics[width=\columnwidth]{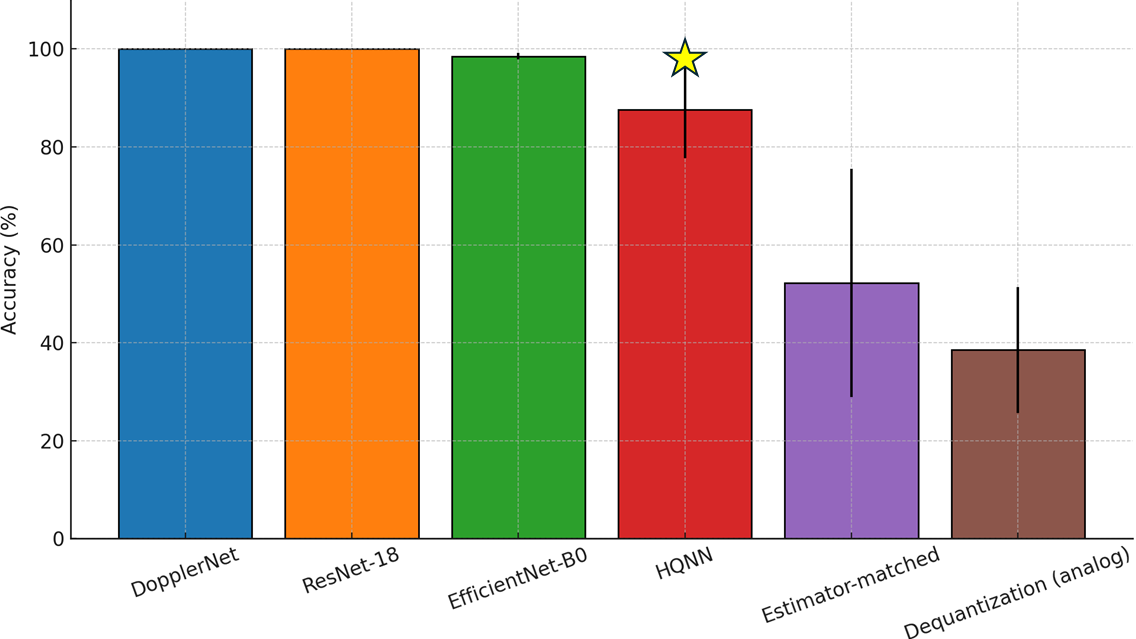}
  \caption{Multi-seed clean-test accuracy on synthetic data (mean $\pm$95\% $t$-CI across five seeds). 
  CNN baselines and HQNN all exceed 95\% accuracy, while attribution controls underperform.}
  \label{fig:multi_seed_acc_synth}
\end{figure}

To ensure the findings from the single run are generalizable, we repeated training with five random seeds (11, 22, 33, 42, 55). Fig.~\ref{fig:multi_seed_acc_synth} reports the mean test accuracy on clean data with 95\% confidence intervals. The results confirm that on average, both CNNs and the HQNN consistently achieve high accuracy, outperforming the attribution controls. Since the performance of the representative run falls within these confidence intervals, our conclusions are statistically robust.

\subsection[Real-Data Training Results (representative run ★)]%
{Real-Data Training Results (representative run \texorpdfstring{$\bigstar$}{*})}
\label{subsec:real-results}

Models are trained on the real dataset for 25 epochs. In a representative held-out split, all CNN baselines exceed $96\%$ accuracy, with DopplerNet and ResNet-18 reaching near-perfect performance ($\approx 99.5\%$). The HQNN achieves a slightly lower test accuracy of $97.0\%$, but remains within 2--3\% of the strongest CNNs. Classical overhead baselines underperform substantially relative to CNNs and HQNN, consistent with their limited \textit{representational capacity}, as their shallow architectures and low parameter counts restrict their ability to capture the task structure. These results confirm that both conventional CNNs and the hybrid HQNN are sufficiently expressive for the real domain under clean conditions.

\begin{table}[!t]
  \caption{Real data: clean-set accuracy on the held-out split (representative run).}
  \label{tab:real_clean}
  \centering\footnotesize
  \begin{tabular}{lcc}
    \hline
    Model                           & Test Accuracy & Train Accuracy \\
    \hline
    DopplerNet                      & 99.5\% & 99.4\% \\
    ResNet-18                       & 99.5\% & 100.0\% \\
    EfficientNet-B0                 & 96.0\% & 99.6\% \\
    HQNN                            & 97.0\% & 95.6\% \\
    Estimator-matched               & 68.5\% & 69.0\% \\
    Dequantization (analog)         & 31.5\% & 34.0\% \\
    \hline
  \end{tabular}
\end{table}

\begin{figure*}[!t]
  \centering
  \subfloat[One person]{%
    \includegraphics[width=0.9\textwidth]{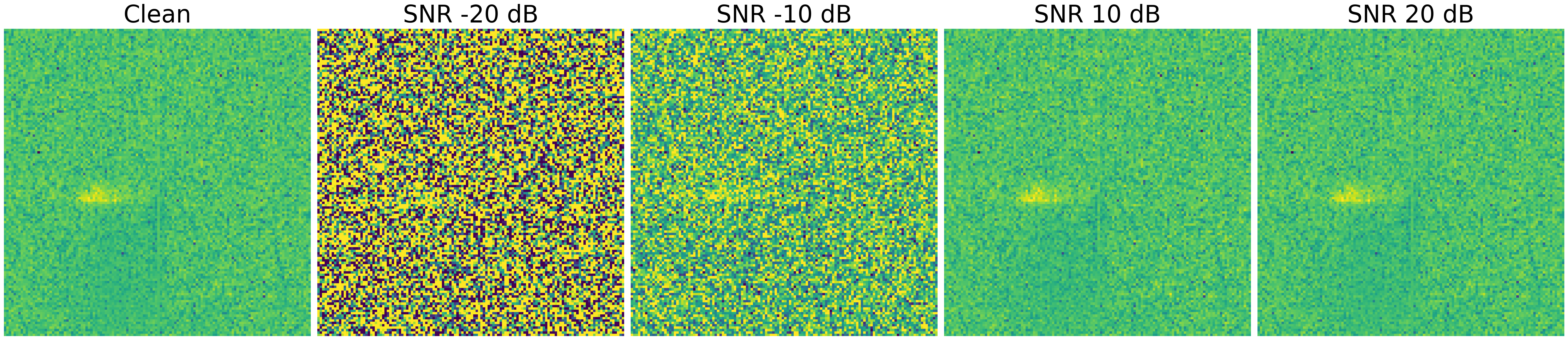}
  }\\[6pt]
  \subfloat[Two people]{%
    \includegraphics[width=0.9\textwidth]{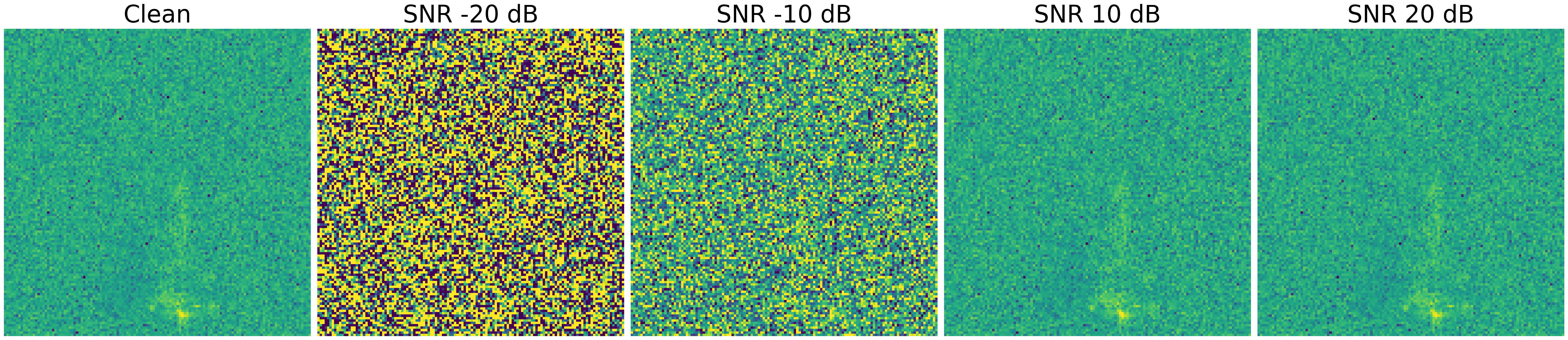}
  }\\[6pt]
  \subfloat[Empty]{%
    \includegraphics[width=0.9\textwidth]{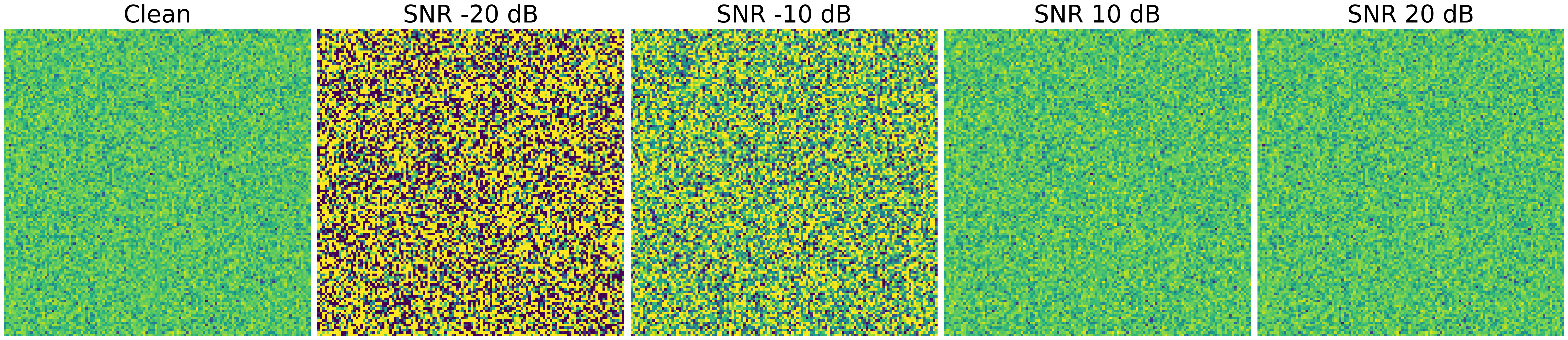}
  }
\caption{Real data (representative run): range--Doppler maps under AWGN injection at different SNRs.}
  \label{fig:real_rdm_snr}
\end{figure*}

\paragraph*{Noise sensitivity using the real dataset}
We next examine model behavior under additive white Gaussian noise (AWGN), injected at test time per~\eqref{eq:awgn} across SNRs $\{-20,-10,+10,+20\}$\,dB. Performance is summarized via Accuracy, Balanced Accuracy (BA), and Macro-F1. Representative range--Doppler maps are shown in Fig.~\ref{fig:real_rdm_snr}: at $-20$ and $-10$\,dB reflections are obscured and class structure vanishes, while at positive SNRs motion streaks and clutter patterns re-emerge, enabling recovery.

\begin{figure*}[!t]
  \centering
  \subfloat[]{%
    \includegraphics[width=0.7\textwidth]{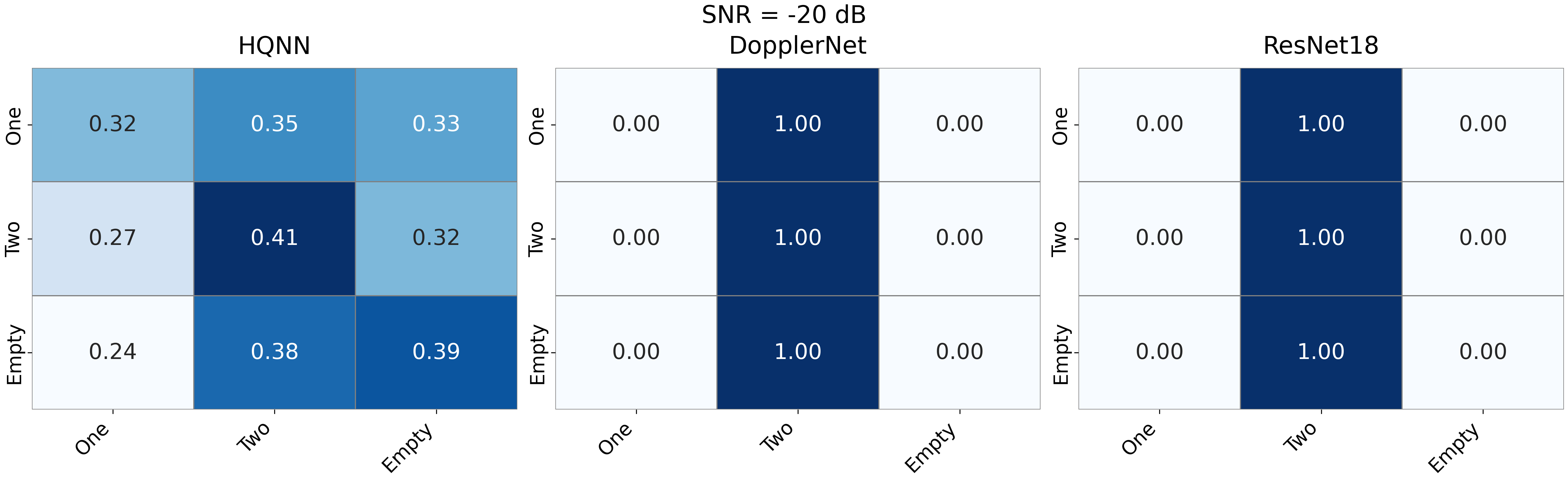}
  }\\[10pt]
  \subfloat[]{%
    \includegraphics[width=0.7\textwidth]{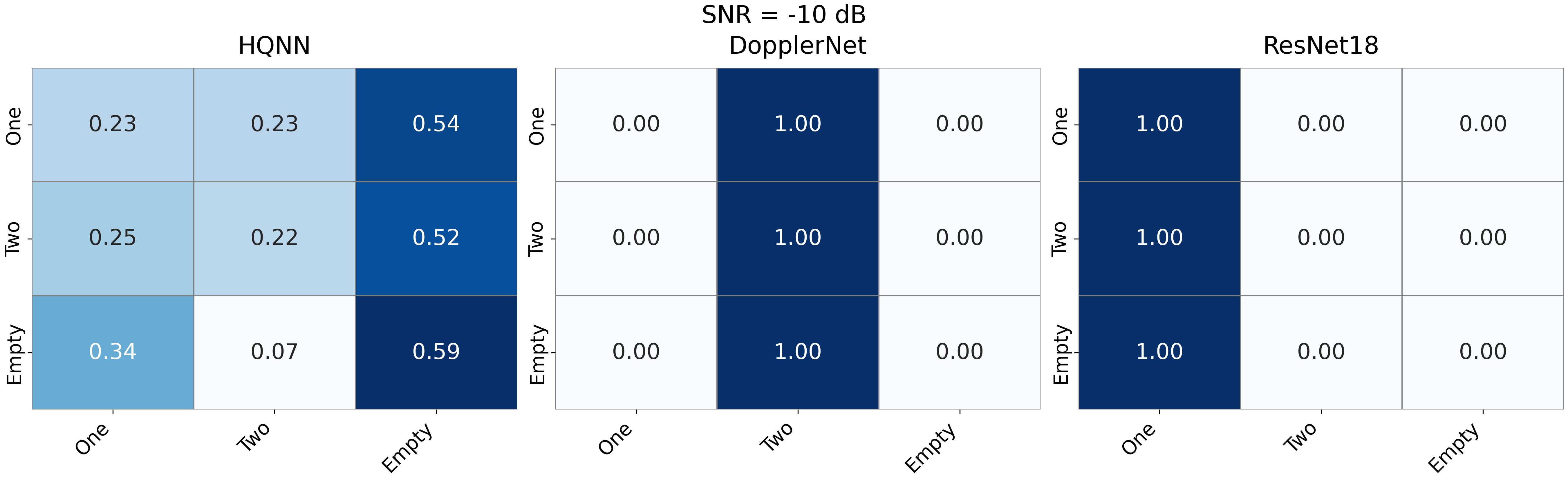}
  }\\[10pt]
  \subfloat[]{%
    \includegraphics[width=0.7\textwidth]{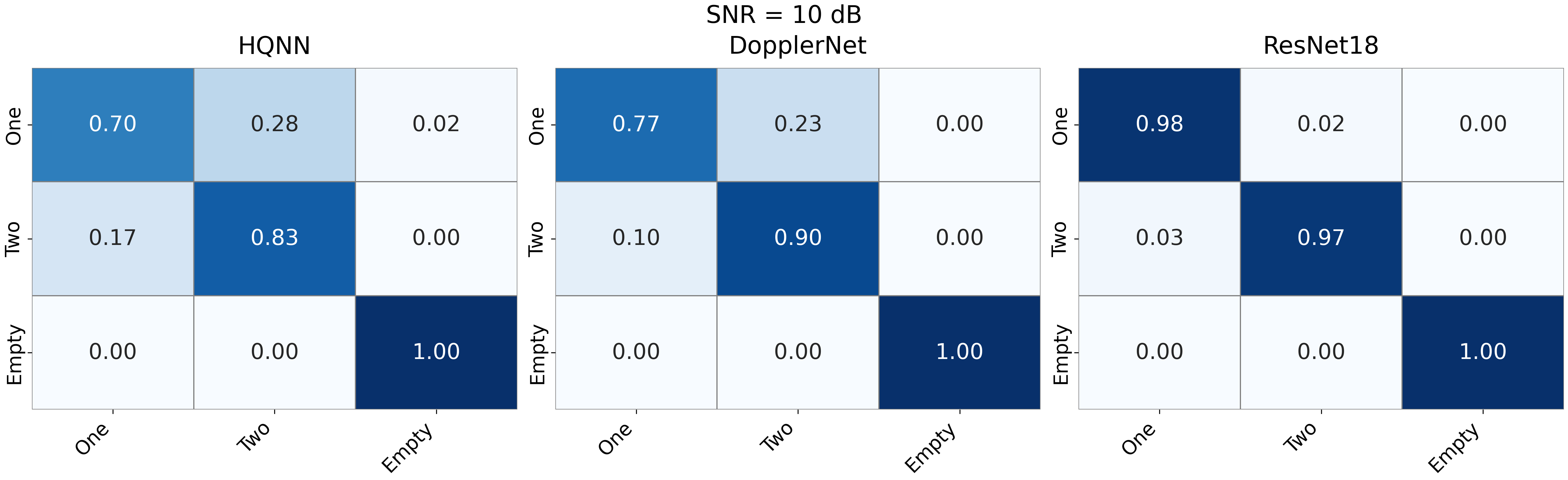}
  }\\[10pt]
  \subfloat[]{%
    \includegraphics[width=0.7\textwidth]{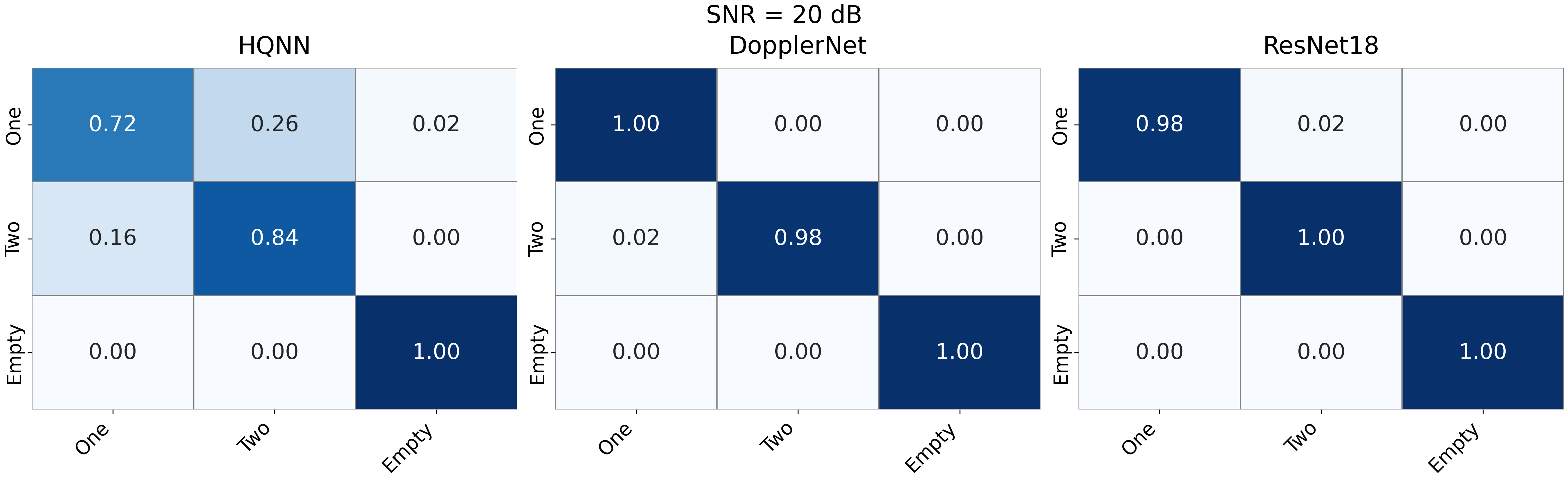}
  }
\caption{Real data (representative run): normalized confusion matrices for HQNN, DopplerNet, and ResNet-18 under different SNRs.}
  \label{fig:real_cm}
\end{figure*}

The models' classification performance under noise is detailed in the confusion matrices in Fig.~\ref{fig:real_cm}. At the lowest SNR of $-20$\,dB, all models degrade to chance-level performance, with DopplerNet exhibiting a strong bias toward the empty class. As SNR improves to $-10$\,dB, HQNN shows early resilience by retaining partial recall for occupied scenes, while the CNNs remain collapsed, predicting only a single class. A dramatic shift occurs at $+10$\,dB, where the CNNs recover sharply and establish dominance. ResNet-18 achieves near-perfect class separation ($\mathrm{BA}\!\approx\!0.984$), decisively outperforming HQNN ($\mathrm{BA}\!\approx\!0.842$). This trend continues at $+20$\,dB, where the strongest CNNs approach 99.5\% accuracy, EfficientNet-B0 follows at $\approx 94.5\%$, and HQNN trails at $\approx 87\%$.

\begin{figure*}[!t]
  \centering
  \includegraphics[width=0.98\textwidth]{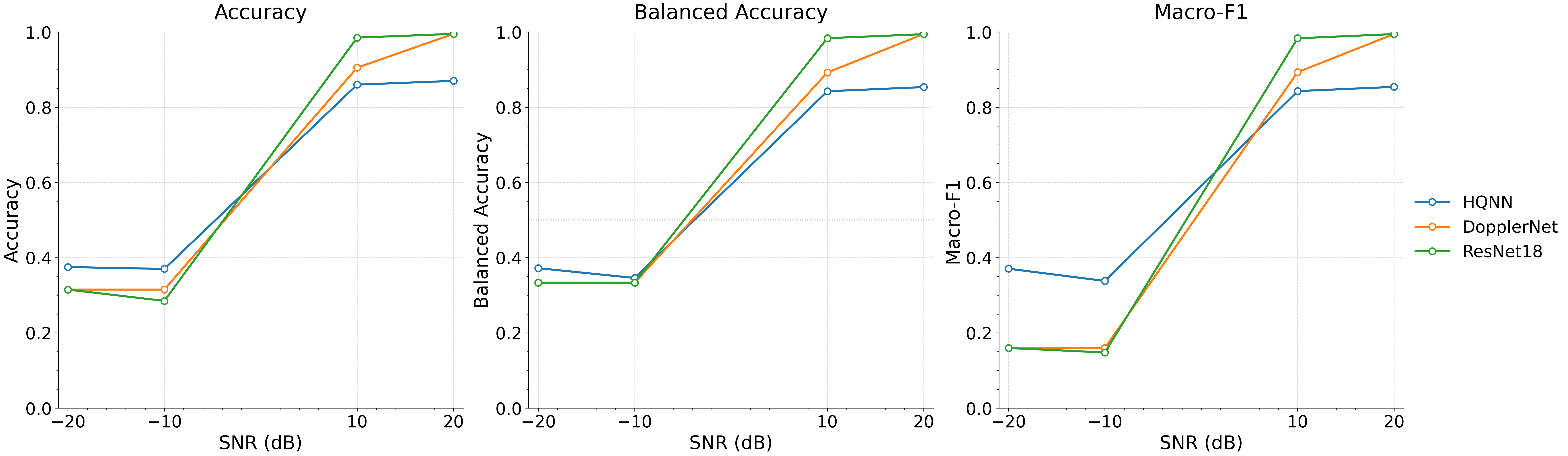}
  \caption{Real data (representative run): test performance with injected AWGN, showing Accuracy, Balanced Accuracy, and Macro-F1 vs.\ SNR.}
  \label{fig:metrics_vs_snr_real}
\end{figure*}

Overall trends are summarized in Fig.~\ref{fig:metrics_vs_snr_real} and Table~\ref{tab:real_results}. BA values near $33\%$ at $-20$ and $-10$\,dB reflect chance-level predictions for all models, but HQNN transitions earlier into recovery. At $+10$\,dB and above, CNNs dominate with steeper recovery curves, whereas HQNN stabilizes at a lower plateau. These results illustrate that HQNN preserves some cues at very low SNRs, but CNNs leverage their greater representational capacity to dominate once signal quality improves.  
(Results in this subsection are from a representative run; multi-seed validation follows below.)

\begin{table*}[!t]
  \caption{Real data (representative run): Balanced Accuracy (BA), Macro-F1, and Accuracy across SNRs for held-out samples with injected AWGN.}
  \label{tab:real_results}
  \centering
  \footnotesize
  \setlength{\tabcolsep}{6pt}
  \renewcommand{\arraystretch}{1.2}
  \begin{tabular}{l|ccc|ccc|ccc|ccc}
    \hline
    \multirow{2}{*}{Model}
      & \multicolumn{3}{c|}{$-20$ dB}
      & \multicolumn{3}{c|}{$-10$ dB}
      & \multicolumn{3}{c|}{$+10$ dB}
      & \multicolumn{3}{c}{$+20$ dB} \\
    \cline{2-13}
      & BA & F1 & Acc & BA & F1 & Acc & BA & F1 & Acc & BA & F1 & Acc \\
    \hline
    DopplerNet
      & 0.333 & 0.160 & 0.315
      & 0.333 & 0.160 & 0.315
      & 0.892 & 0.893 & 0.905
      & 0.995 & 0.994 & 0.995 \\
    ResNet-18
      & 0.333 & 0.160 & 0.315
      & 0.333 & 0.148 & 0.285
      & 0.984 & 0.983 & 0.985
      & 0.994 & 0.994 & 0.995 \\
    EfficientNet-B0
      & 0.319 & 0.182 & 0.300
      & 0.317 & 0.242 & 0.290
      & 0.934 & 0.933 & 0.940
      & 0.940 & 0.939 & 0.945 \\
    HQNN
      & 0.372 & 0.371 & 0.375
      & 0.346 & 0.338 & 0.370
      & 0.842 & 0.843 & 0.860
      & 0.854 & 0.854 & 0.870 \\
    \hline
  \end{tabular}
\end{table*}

\subsubsection{Multi-seed analysis}
\begin{figure}[!t]
  \centering
  \includegraphics[width=\columnwidth]{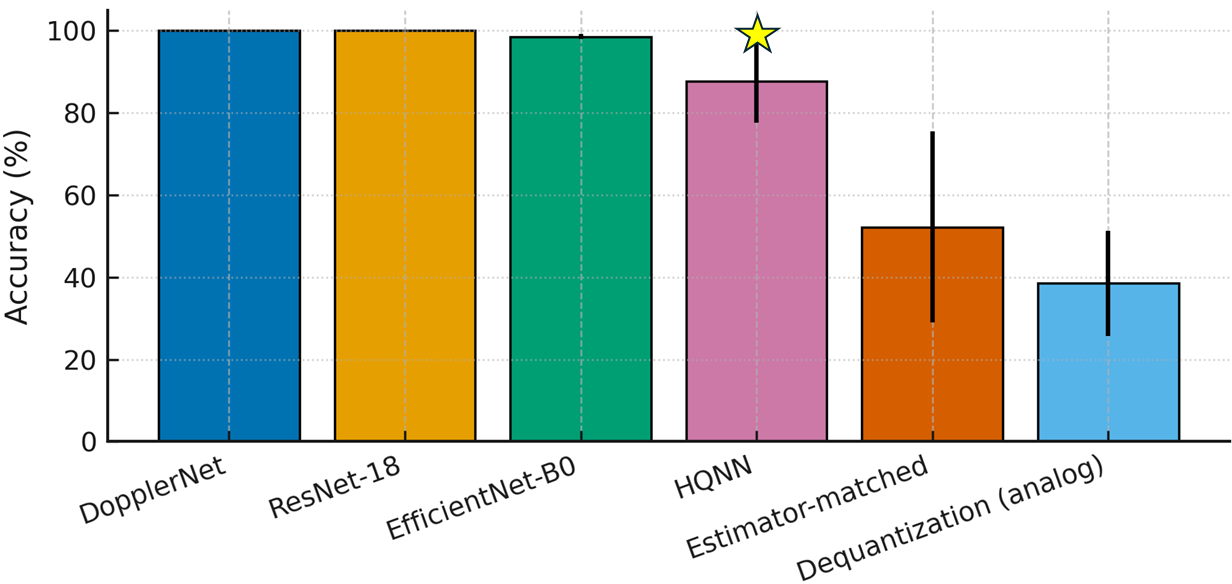}
  \caption{Multi-seed clean-test accuracy on real data (mean $\pm$ 95\% $t$-CI across five seeds). 
  CNN baselines remain most accurate and consistent; HQNN exceeds parameter-matched classical heads but does not surpass CNNs.}
  \label{fig:multi_seed_acc}
\end{figure}

To validate our single-run findings, we repeated the training on the real dataset with five distinct random seeds. As shown in Fig.~\ref{fig:multi_seed_acc}, the CNN baselines remained the most accurate and consistent models across all runs. While the HQNN did not surpass the CNNs, it consistently outperformed the parameter-matched classical heads. Since the performance differences from the representative run fall within these confidence intervals, our conclusions regarding the clean and noisy trends in Tables~\ref{tab:real_clean}--\ref{tab:real_results} are statistically robust.

\subsubsection*{Note on Computational Cost}
For completeness, we briefly address the computational cost of training. Across both the synthetic and real datasets, the HQNN shows a substantial training-time overhead (on the order of $47\times$ vs.\ CNNs) despite its much smaller parameter count. This overhead is characteristic of classically simulating parameterized quantum circuits and arises from PQC evaluation, limited batching, and software latency. While such costs are a known bottleneck of simulation, our focus in this work is on representational efficiency and robustness under noise rather than claims of computational speedup. A fuller discussion of runtime constraints and their implications for quantum hardware deployment is deferred to Sec.~\ref{sec:discussion}.

\subsection{Ablation on Training Fraction (Sample Efficiency)}
To evaluate sample efficiency, we trained all models on random subsamples of the training set (10--50\%) using a fixed configuration (identical splits, optimizer, and epochs). This experiment, conducted on noise-free data, assesses how each architecture scales when training data are limited. Evaluation is always performed on the full held-out set. Results are summarized in Tables~\ref{tab:ablation_real} and~\ref{tab:ablation_synth}.

On real data, deeper convolutional networks demonstrated clear advantages in low-data regimes. With only 10\% of the training set, EfficientNet-B0 achieved 0.79 BA and ResNet-18 reached 0.66, while HQNN lagged at 0.63 and DopplerNet collapsed to 0.38. As the training fraction increased, the gap widened: at 50\% of the data, CNNs approached saturation (BA $\approx$0.99), whereas HQNN improved modestly to 0.75, trailing all CNN baselines. These results indicate that when labeled data are scarce, deeper architectures exploit redundancy in range–Doppler maps more effectively to generalize from limited examples.

On synthetic data, the pattern was less pronounced. HQNN improved more rapidly, reaching 0.82 BA at 30\% of the training set, close to DopplerNet (0.85) and EfficientNet-B0 (0.83). By 50\%, all models converged to a narrow range of 0.85--0.87 BA, with ResNet-18 holding a slight lead. This reflects the idealized nature of the digital twin, where reduced variability diminishes architectural differences.

Taken together, these findings show that HQNN is parameter-efficient but not sample-efficient. In real-world data, CNNs retain a clear advantage in generalization, whereas in controlled synthetic domains the gap narrows. This suggests that compact hybrid models may benefit most from refinements such as radar-aware embeddings or shallow but wider parameterized circuits if they are to compete in data-limited sensing scenarios.

\begin{table}[H]
  \centering
  \caption{Balanced Accuracy (BA) across training fractions for real data.}
  \label{tab:ablation_real}
  \footnotesize
  \setlength{\tabcolsep}{6pt}
  \renewcommand{\arraystretch}{1.2}
  \begin{tabular}{lccc}
    \hline
    Model & 10\% & 30\% & 50\% \\
    \hline
    HQNN            & 0.63 & 0.67 & 0.75 \\
    DopplerNet      & 0.38 & 0.93 & 0.93 \\
    ResNet-18       & 0.66 & 0.83 & 0.89 \\
    EfficientNet-B0 & 0.79 & 0.94 & 0.99 \\
    \hline
  \end{tabular}
\end{table}

\begin{table}[H]
  \centering
  \caption{Balanced Accuracy (BA) across training fractions for synthetic data.}
  \label{tab:ablation_synth}
  \footnotesize
  \setlength{\tabcolsep}{6pt}
  \renewcommand{\arraystretch}{1.2}
  \begin{tabular}{lccc}
    \hline
    Model & 10\% & 30\% & 50\% \\
    \hline
    HQNN            & 0.47 & 0.82 & 0.85 \\
    DopplerNet      & 0.76 & 0.85 & 0.86 \\
    ResNet-18       & 0.75 & 0.83 & 0.87 \\
    EfficientNet-B0 & 0.73 & 0.83 & 0.86 \\
    \hline
  \end{tabular}
\end{table}

\section{Discussion}
\label{sec:discussion}

\paragraph*{Clean baselines establish solvability, not differentiation.}
Across both domains, three-class occupancy is not intrinsically difficult under noise-free conditions: all CNN baselines exceed $96\%$ test accuracy, and HQNN converges near $97\%$. Synthetic data even allows near-perfect scores across all models. Control heads without the PQC collapse below $75\%$, suggesting that the quantum layer contributes nontrivial capacity beyond the 2-D bottleneck. Concretely, the dequantized heads reach only $71.7\%$ and $60.2\%$ on synthetic data (Table~\ref{tab:synth_clean}) and $68.5\%$ and $31.5\%$ on real data (Table~\ref{tab:real_clean}), despite sharing the same CNN backbone and bottleneck. This distinction is notable because the PQC introduces only four trainable rotation angles across two qubits; the gain thus arises from its entangling structure rather than added parameter count. These results establish that the task is solvable, but they do not distinguish architectures; informative differences emerge only once models are stressed.

\paragraph*{Noise sensitivity reveals architectural biases.}
Injecting noise highlights the different architectural tendencies of the models. In the synthetic domain, the HQNN begins recovering at $+10$\,dB (BA $0.792$ vs.\ ResNet-18 at $0.755$; Table~\ref{tab:synth_results}). In the real domain, by contrast, the CNNs recover more sharply: at $+10$\,dB, ResNet-18 exceeds HQNN by $\Delta$BA $\approx 0.142$ (0.984 vs.\ 0.842), and at $+20$\,dB by $\approx 0.140$ (0.994 vs.\ 0.854) (Table~\ref{tab:real_results}). This contrast—earlier onset for HQNN in synthetic data, steeper recovery for CNNs in real data—illustrates how each architecture relies on distinct feature representations. In clean synthetic trials, both ResNet-18 and HQNN approach $98\%$ BA, whereas in real data the higher capacity of the CNNs leads to stronger performance. All PQC evaluations used 4096-shot sampling, introducing hardware-like measurement stochasticity; this may contribute to broader intervals for HQNN at low SNRs (multi-seed CIs shown in Figs.~\ref{fig:multi_seed_acc_synth},~\ref{fig:multi_seed_acc}).

\paragraph*{Digital twin as controlled instrument.}
The geometric simulator provides more than training data; it serves as a structured instrument for probing sensitivity and collapse modes. Because models are trained and evaluated independently within each domain, comparisons across synthetic and real are best interpreted as controlled analogies rather than literal DT-to-real transfer. The consistent mid-SNR behavior observed in both domains, mirrored only qualitatively in real data—where HQNN shows early consistency rather than performance gains—suggests that the DT captures essential structural aspects of the sensing problem, including multipath and clutter arising from scene geometry. At the same time, its limitations are clear: absolute power levels are mismatched by approximately $170$\,dB, oscillator phase noise and receiver impairments are not modeled, and scene layouts remain static. These idealizations define a concrete roadmap for advancing DT fidelity: calibrating mean/variance statistics to $60$\,GHz hardware, incorporating hardware-specific noise sources and more stochastic clutter models, and randomizing geometry and pose. Such refinements will strengthen the DT not only as a data generator but also as a predictive testbed for hybrid architectures~\cite{Chipengo2021HighFid}.

\paragraph*{Sample efficiency remains limited for our specific data processing case.}
The ablation on training fractions highlights a central limitation of the present HQNN design: it fails to match CNNs in data-scarce regimes. On real data, HQNN lags behind by $20$--$27$ BA points at $30\%$ of the training set (HQNN $0.67$ vs.\ ResNet-18 $0.83$ and EfficientNet-B0 $0.94$; Table~\ref{tab:ablation_real}), only converging toward CNN performance when more data are available. This trade-off between parameter compactness and generalization is particularly consequential in radar applications, where labeled real-world datasets are costly and time-intensive to collect. The gap is consistent with the HQNN's architecture: a radar-agnostic 2-D latent passed to a two-qubit PQC that is compact and stable, but under-expressive in width. By contrast, CNNs exploit convolutional hierarchies to capture structural redundancy in RDMs, enabling strong performance even with limited training samples. Thus, the HQNN sacrifices generalization when data are scarce in exchange for compactness and mid-SNR robustness. Promising directions to address this include: (i) incorporating radar-informed feature maps into the quantum layer, (ii) exploring wider but shallow PQCs that expand representational capacity without prohibitive depth, and (iii) repositioning the PQC earlier in the pipeline, where it could contribute directly to feature extraction rather than only serving as a compact classifier head. Such refinements will be essential if HQNNs are to become competitive in few-shot or data-limited sensing scenarios—precisely where their parameter efficiency has the greatest potential impact~\cite{Grossi2022TQE}.

\paragraph*{Stability, efficiency, and architectural trade-offs.}
Our multi-seed analysis confirms the statistical robustness of these findings. Across five independent seeds, the performance hierarchy was stable: CNNs consistently led, while HQNN reliably outperformed the attribution controls (Figs.~\ref{fig:multi_seed_acc_synth},~\ref{fig:multi_seed_acc}). This stability is linked to a core trade-off between parameter efficiency and computational cost. The HQNN is exceptionally compact ($\sim6.6\times10^4$ parameters), yet in our simulations its training is $\sim$47$\times$ slower than the CNNs on average. The source of this overhead is the gradient computation method. A CNN calculates gradients for all its parameters in a single backward pass. In contrast, differentiating the PQC with the parameter-shift rule requires numerous circuit evaluations for each trainable angle. In our implementation, this amounts to 416 PQC evaluations per batch, and with $\sim$66 optimization steps per epoch over 15 epochs, a single run exceeds $4\times10^5$ circuit evaluations. On native quantum hardware, these would correspond to physical circuit executions rather than costly statevector simulations, potentially eliminating much of this burden. The key insight is that in the current paradigm, the HQNN's value is not runtime efficiency but its compact inductive bias, which produces its distinct stability patterns across seeds and mid-SNR conditions. Benchmarking on native QPUs will be necessary to assess training latency and energy cost relative to GPU-based CNNs under identical datasets and splits.

\paragraph*{Performance in context: early-stage quantum hybrids and established CNNs.}
The comparison presented here must be interpreted in light of the differing levels of technological maturity. CNNs such as ResNet represent more than a decade of architectural refinement and optimization for classical hardware (e.g., GPUs), whereas hybrid quantum--classical models remain at an early stage of development. Within this study, the HQNN achieves competitive results in certain synthetic regimes despite its minimal design (two qubits; four trainable angles) and nascent toolchain. This should not be viewed as a direct shortcoming relative to CNNs, but rather as an encouraging sign that compact quantum-informed models can already approach the performance of established baselines in specific conditions. As quantum hardware, algorithms, and software frameworks evolve, it is reasonable to expect the gap between these approaches to shift in meaningful ways for relevant sensing tasks.

\paragraph*{Limitations and forward directions.}
The present results remain simulator-based and use a classical backend only. Zero-shot transfer highlights that additive noise is not the sole challenge: real-world mismatches include absolute power and noise-floor calibration (an offset of approximately $170$\,dB in our digital twin), hardware-specific impairments such as phase noise and nonlinearities, and stochastic clutter not represented in the current pipeline. While SBR captures multipath and deterministic clutter imposed by scene geometry, it does not reproduce dynamic or environment-specific clutter distributions that dominate real deployments. Bridging this gap requires co-design: advancing DT fidelity through calibrated power levels, realistic noise-floor models, and dynamic scene variability, while in parallel incorporating radar-specific embeddings into quantum layers and benchmarking against noise-resistant CNNs. On the quantum side, validation on NISQ hardware will be essential to translate parameter efficiency into practical utility, and to assess whether hardware noise undermines or reinforces the stability trends observed here~\cite{wang2021noise}. For healthcare applications, coupling radar counting with fall-detection pipelines raises further challenges of alarm fatigue and contextual inference, pointing to the value of hybrid approaches that are compact, interpretable, and generalizable across environments~\cite{Du2020GRSL}.

\section{Conclusion}
\label{sec:conclusion}
This work presents a compact two-qubit Hybrid Quantum Neural Network designed for radar-based occupancy classification and its first side-by-side comparison with established CNN architectures. Using a physics-informed $60\,\mathrm{GHz}$ digital twin and real measurements under matched training conditions, the study reveals a domain-dependent trade-off. In the real domain, CNNs achieve higher accuracy and stronger sample efficiency. In the synthetic domain, the HQNN attains comparable performance with up to $170\times$ fewer parameters and begins recovery earlier in mid-SNR conditions.
A central finding comes from dequantized control experiments in both domains. Replacing the quantum layer with a parameter-matched classical head causes a sharp performance collapse, confirming that the parameterized quantum circuit provides essential representational capacity rather than acting as a minimal bottleneck. This shows that even compact hybrid quantum models can contribute meaningfully to feature representation in radar perception.
Building on these findings, a controlled hardware study is necessary to assess computational efficiency. HQNN and CNN models should be trained on identical data and compared on quantum and classical processors under matched noise conditions, evaluating accuracy, time to target, latency, and energy consumption.
In parallel, sim-to-real calibration of the digital twin is required to align simulated and measured radar statistics, improving benchmarking fidelity and enabling the twin to serve as a predictive testbed for future architectures.
Future work will systematically expand the quantum design space to determine where it provides the most value: key experiments will include varying PQC placement across network stages, scaling qubit count and gate depth to map the expressivity-versus-noise trade-off, testing alternative data encodings, and exploring compact all-quantum pipelines to bound the potential of this approach.
\section*{Acknowledgments}
The authors would like to acknowledge NSERC, MITACS, and Synopsys for their support. We would like to acknowledge CMC Microsystems, manager of the FABrIC project funded by the Government of Canada, for the provision of products and services that facilitated this research.

\bibliographystyle{IEEEtran}
\bibliography{clean_reference}

\vspace{-30pt}
\begin{IEEEbiography}[{\includegraphics[width=1in,height=1.25in,clip,keepaspectratio]{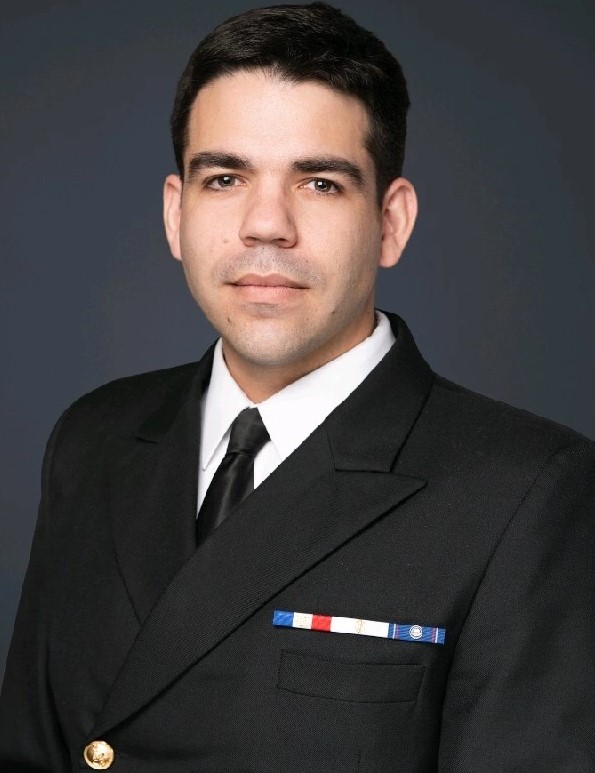}}]{Sebastian Ratto Valderrama}
(Student Member, IEEE) is a Chilean naval officer with 12 years of service and a Ph.D. student in Electrical and Computer Engineering at the University of Waterloo, where he is affiliated with the Wireless Sensors and Devices Laboratory and the Advanced Concepts Research Laboratory. He received his B.Sc. and B.Eng. degrees in Naval Electronics Engineering from the Academia Polit\'ecnica Naval, Vi\~na del Mar, Chile, in 2022. From 2022 to 2024, he participated in ship-system modernization projects aboard LAM-31 \emph{Chipana} in collaboration with industry partners. At Waterloo, he chairs the IEEE MTT-S Student Branch Chapter. His doctoral research, under the supervision of Dr. George Shaker and Dr. Omar Ramahi, focuses on mmWave radar sensing, digital-twin modeling for healthcare, and hybrid quantum--classical machine learning for radar signal processing.
\end{IEEEbiography}
\vspace{-30pt}
\begin{IEEEbiography}[{\includegraphics[width=1in,height=1.25in,clip,keepaspectratio]{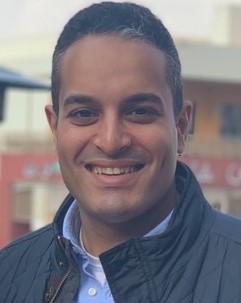}}]{Ahmed N. Sayed}
(Senior Member, IEEE) received the B.Sc. and M.Sc. degrees in electrical engineering from the Military Technical College, Cairo, Egypt, in 2009 and 2015, respectively, and the Ph.D. degree in electrical and computer engineering from the University of Waterloo, Waterloo, ON, Canada, in 2024. His research interests include radar detection, digital and statistical signal processing, and machine learning for autonomous sensing and surveillance, with emphasis on mmWave systems, tracking, and robust perception.
\end{IEEEbiography}
\begin{IEEEbiography}[{\includegraphics[width=1in,height=1.25in,clip,keepaspectratio]{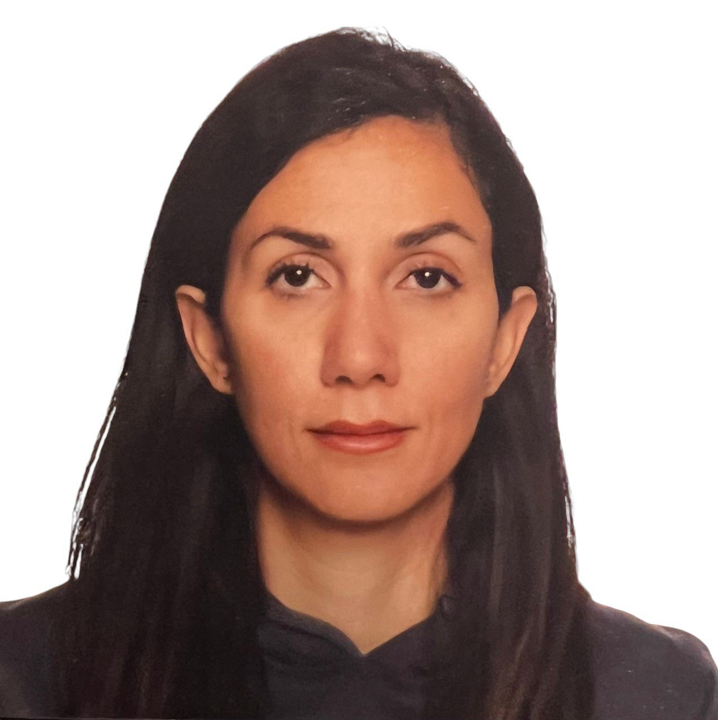}}]{Neda Rojhani}
(Senior Member, IEEE) received her Ph.D. in Electronics and Electromagnetism Engineering from the University of Florence, Italy, in 2019. She is currently with the Department of Electrical and Computer Engineering at the University of Waterloo, Canada, specializing in advanced radar signal processing, including MIMO radar, SAR, InSAR, GPR, compressive sensing, the Cram\'er--Rao bound, and antenna design. Her research extends to AI, machine learning, and optimization algorithms to enhance emerging technologies. She is an active member of the TC-27 Connected and Autonomous Systems Committee (IEEE MTT-S), and she contributes to advancements in connected and autonomous systems.
\end{IEEEbiography}

\begin{IEEEbiography}[{\includegraphics[width=1in,height=1.25in,clip,keepaspectratio]{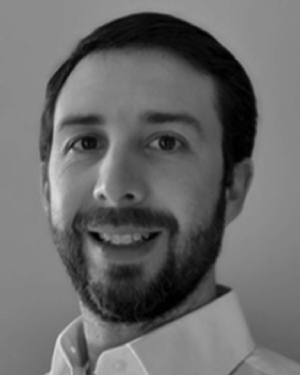}}]{Arien P. Sligar}
(Member, IEEE) received the B.S. and M.Sc. degrees in electrical engineering from Oregon State University, Corvallis, OR, USA, in 2004 and 2006, respectively, with a focus on electromagnetics and microwave components. Since 2006, he has been with Ansys, Inc., where he is a Principal Engineer focusing on advanced applications of numerical simulation for electromagnetics and electronics. He works with leading technology companies on the design of antennas and complex antenna systems, microwave components, and high-speed electronics, and on deploying automated simulation workflows for challenging engineering problems. His interests include electromagnetic field simulation, modeling of complex RF systems, and design automation.
\end{IEEEbiography}
\begin{IEEEbiography}[{\includegraphics[width=1in,height=1.25in,clip,keepaspectratio]{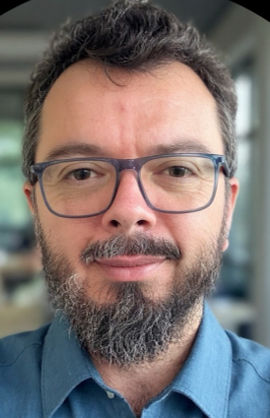}}]{Jose R. Rosas\mbox{-}Bustos}
is a Ph.D. candidate at the University of Waterloo, Waterloo, ON, Canada, where his research focuses on quantum-safe cryptography, quantum communication, and quantum information processing. He is also the Science Director of the AQT Institute and the CEO of EigenQ, a quantum-technology company focused on quantum-resistant technologies, quantum communication, and quantum information processing. His interests include quantum-resistant cryptography, quantum systems architectures, and the transition of laboratory prototypes into deployable technologies.
\end{IEEEbiography}

\begin{IEEEbiography}[{\includegraphics[width=1in,height=1.25in,clip,keepaspectratio]{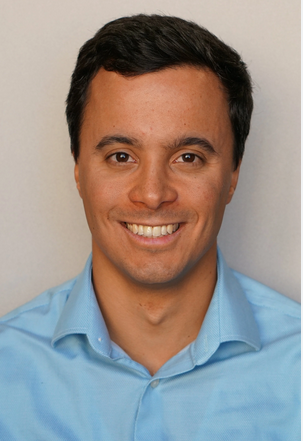}}]{Luke C.\ G.\ Govia}
(Member, IEEE) is the Manager for Quantum Technologies at CMC Microsystems, Waterloo, ON, Canada. A theoretical physicist by training, his expertise includes superconducting devices, open quantum systems, quantum error mitigation and correction, characterization and benchmarking of quantum computers, and neuromorphic computing. He received the Ph.D. degree in physics from Saarbr\"ucken University, Germany, and the M.Sc. and B.Sc. degrees in physics from the University of Waterloo, Waterloo, ON, Canada. He serves on the Unitary Foundation's Open Quantum Benchmark Committee and was Secretary--Treasurer (2021--2023) of the American Physical Society's Division of Quantum Information.
\end{IEEEbiography}

\begin{IEEEbiography}[{\includegraphics[width=1in,height=1.25in,clip,keepaspectratio]{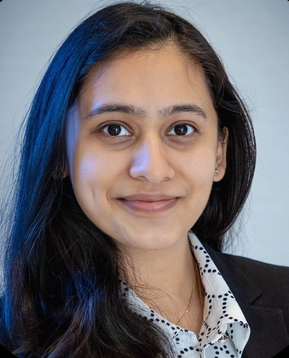}}]{Saasha Joshi}
(Member, IEEE) is a Staff Scientist in Quantum Computing at CMC Microsystems, Canada. Her expertise includes quantum machine learning, software development, and optimization. She actively contributes to open-source quantum software and engages in STEM education initiatives. She holds the M.Sc. degree in Computer Science from the University of Victoria, Canada, and the B.E. degree in Computer Science Engineering from Panjab University, India.
\end{IEEEbiography}
\begin{IEEEbiography}[{\includegraphics[width=1in,height=1.25in,clip,keepaspectratio]{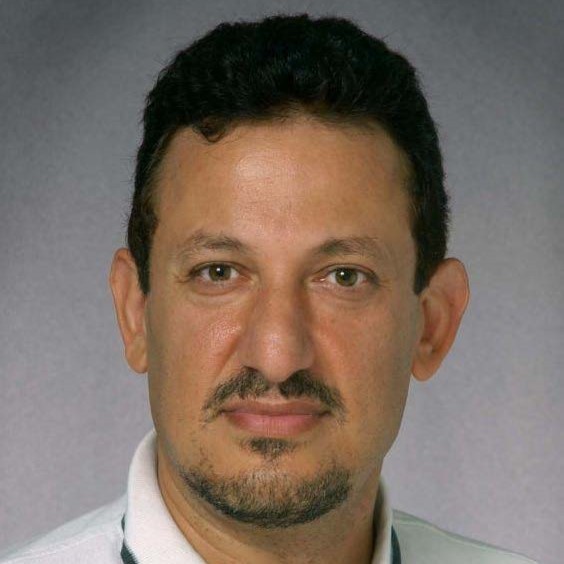}}]{Omar M. Ramahi}
(Fellow, IEEE) was born in Jerusalem, Palestine. He received the B.S. degree (Hons.) in mathematics and electrical and computer engineering from Oregon State University, Corvallis, OR, USA, in 1984, and the M.S. and Ph.D. degrees in electrical and computer engineering from the University of Illinois at Urbana--Champaign, Champaign, IL, USA, in 1986 and 1990, respectively. He was with Digital Equipment Corporation (currently HP), Maynard, MA, USA, where he was a member of the Alpha Server Product Development Group. In 2000, he joined the James Clark School of Engineering, University of Maryland, College Park, MD, USA, as an Assistant Professor and later as a tenured Associate Professor, where he was also a Faculty Member with the CALCE Electronic Products and Systems Center. He is currently a Professor with the Department of Electrical and Computer Engineering, University of Waterloo, ON, Canada. He has authored or co-authored over 500 journal articles and conference technical articles on topics related to electromagnetic phenomena and computational techniques. He has co-authored the book EMI/EMC Computational Modeling Handbook (first edition: Kluwer, 1998, second edition: Springer-Verlag, 2001. Japanese edition published in 2005). Prof. Ramahi received the 2004 University of Maryland Pi Tau Sigma Purple Cam Shaft Award, the Excellent Paper Award from the 2004 International Symposium on Electromagnetic Compatibility, Sendai, Japan, the 2010 University of Waterloo Award for Excellence in Graduate Supervision, the IEEE EMC Society Technical Achievement Award in 2012, and the 2022 University of Waterloo Engineering Research Excellence Award.
\end{IEEEbiography}
 
\begin{IEEEbiography}[{%
  \protect\colorbox{white}{%
    \includegraphics[width=1in,height=1.25in,clip,keepaspectratio]{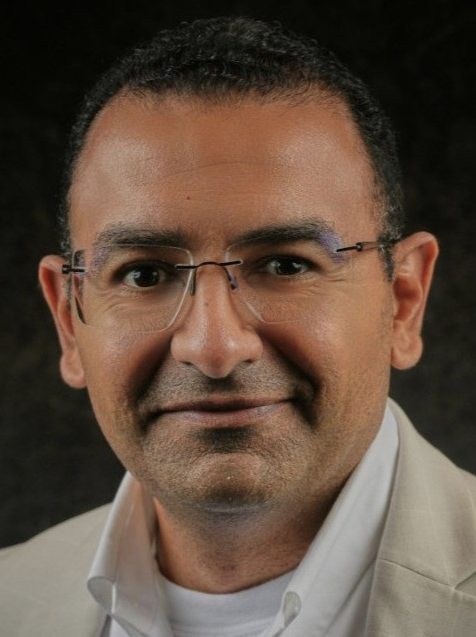}}}]{George Shaker (S'97--M'11--SM'15)}is the lab director of the Wireless Sensors and Devices Laboratory at the University of Waterloo, where he is an adjunct associate professor at the Department of Electrical and Computer Engineering. Previously, he was an NSERC scholar at Georgia Institute of Technology. Dr. Shaker also held multiple roles with RIM's (BlackBerry). He is also the Chief Scientist at Spark Tech Labs, which was co-founded in 2011. With over twenty years of industrial experience in technology, and more than ten years as a faculty member leading project related to the application of wireless sensor systems for healthcare, automotives, and unmanned aerial vehicles, Prof. Shaker has many design contributions in commercial products available from startups and multinationals. A sample list includes Google, COM DEV, Honeywell, Blackberry, Spark Tech Labs, Bionym, Lyngsoe Systems, ON Semiconductors, Ecobee, Medella Health, NERV Technologies, Novela, Thalmic Labs, North, General Dynamics Land Systems, General Motors, Toyota, Maple Lodge Farms, Rogers Communications, and Purolator. He is currently an IEEE AP-S Distinguished Industry Speaker, and IEEE Sensors Council Distinguished Lecturer.

Dr. Shaker has authored/coauthored 200+ publications and 35+ patents/patent applications. George has received multiple recognitions and awards, including the the IEEE AP-S Best Paper Award (the IEEE AP-S Honorable Mention Best Paper Award (4 times to-date), the IEEE Antennas and Propagation Graduate Research Award, the IEEE MTT-S Graduate Fellowship, the Electronic Components and Technology Best of Session paper award, and the IEEE Sensors most popular paper award. Four papers he co-authored in IEEE journals were among the top 25 downloaded papers on IEEEXplore for several consecutive months. He was the supervisor of the student team winning the third best design contest at IEEE AP-S 2016 and 2025, co-author of the ACM MobileHCI 2017 best workshop paper award, and the 2018 Computer Vision Conference Imaging Best Paper Award. He co-received with his students several research recognitions including the NSERC Top Science Research Award 2019, IEEE APS HM paper award (2019, 2022, 2023, 2024, and 2025), Biotec top demo award 2019, arXiv top downloaded paper (medical device category) 2019, Velocity fund 2020, NASA Tech Briefs HM Award (medical device category) 2020, UW Concept 2021, UK Dragons Canadian Competition 2021, CMC Nano 2021, COIL COLAB 2022, Wiley Engineering Reports top downloaded paper for 2022, Canadian Space Agency Cubesat Design winner 2023, IEEE NEMO Best Paper Award 2024, Nature Communications Engineering Top 25 downloaded papers in 2024, IEEE MAPCON Best Paper Award 2024, and iWAT 2025 paper finalist.
\end{IEEEbiography}

\end{document}